\documentclass[12pt,12pt]{article}
\usepackage{graphicx}
\usepackage{epsfig}
\usepackage{amsmath,amssymb}

\catcode`\@=11
\newdimen\ex@
\def\beq{\begin{equation}}
\def\eeq{\end{equation}}
\def\beqa{\begin{eqnarray}}
\def\eeqa{\end{eqnarray}}
\newcommand{\ba}{\begin{eqnarray}}
\newcommand{\ea}{\end{eqnarray}}
\newcommand\BA{\begin{array}}
\newcommand\EA{\end{array}}

\begin{document}
\def\thefootnote{\fnsymbol{footnote}}

\title{\bf
An attempt at a resonating mean-field \\
theoretical description of thermal behavior \\
of two-gap superconductivity}
\vskip0.4cm
\author{Seiya Nishiyama$\!$\footnotemark[1] ,
Jo\~ao da Provid\^{e}ncia$\!$\footnotemark[2]\\
Constan\c{c}a Provid\^{e}ncia\footnotemark[3]~~and
Hiromasa Ohnishi\footnotemark[4]\\
\\[-0.2cm]
Centro de F\'\i sica Computacional,
Departamento de F\'\i sica,\\
Universidade de Coimbra,
P-3004-516 Coimbra, Portugal\\
\\[-0.cm]
Nanosystem Research Institute (NRI),\\
National Institute of Advanced Science and Technology (AIST),\\
1-1-1, Umezono, Tsukuba, Ibaraki 305-8568, Japan\footnotemark[4]\\
\\[-0.2cm]
{\it Dedicated to the Memory of Hideo Fukutome}}

\maketitle
\footnotetext[1]{
Corresponding author. E-mail address: seikoceu@khe.biglobe.ne.jp}
\footnotetext[2]{
E-mail address: providencia@teor.fis.uc.pt}
\footnotetext[3]
{E-mail address: cp@teor.fis.uc.pt}
\footnotetext[4]
{E-mail address: hiro.ohnishi@aist.go.jp}

\vspace{-0.50cm}

%%%%%%%%%%
%                        %
%  0   Abstract    %
%                        %
%%%%%%%%%%

\begin{abstract}
\vskip0.1cm
The resonating mean-field theory (Res-MFT)
has been applied and shown to effectively describe
two-gap superconductivity (SC).
Particularly at $T \!=\! 0$
using a suitable chemical potential,
the two-gap SC in MgB$_2$ has been well described by
the Res-Hartree-Bogoliubov theory (Res-HBT).
The Res-HB ground state generated with HB wave functions
almost exhausts the ground-state correlation energy
in all the correlation regimes.
In this paper
we make an attempt at a Res-MF theoretical description
of thermal behavior of the two-gap SC.
In an {\em equal energy-gap} case
we find a new formula leading to a higher $T_c$ than the $T_c$ of
the usual HB formula.

\vspace{0.2cm}

\end{abstract}

\newpage

%%%%%%%%%%%%
%                              %
%  1  Introduction     %
%                              %
%%%%%%%%%%%%

\def\thesection{\arabic{section}}
\setcounter{equation}{0}
\renewcommand{\theequation}{\arabic{section}.\arabic{equation}}

\section{Introduction}

\vspace{-0.3cm}

~~~~A two-gap superconductivity (SC) of
magnesium diboride $\mbox{MgB}_2$
with critical temperature $T_c \!=\! 39$K
has been discovered nearly a decade ago
\cite{Akimitsu.01}.
Hitherto,
intensive studies had been made to raise the $T_c$
of usual BCS superconductor in the weak coupling regime
\cite{BCS.57,Bogo.59,BTS.59}
and to obtain Eliashberg's critical temperature
in the strong coupling
\cite{Eliashberg.60,Schrieffer.64,GK.82}.
The $T_c \!=\! 39$K
is close to or even above the upper theoretical value
predicted by the BCS theory
\cite{Parks.69}. 
Through {\em ab initio} density functional computations
it has been estimated as 22K by Kortus et al.
\cite{KMBAB.01}.
The existence of two energy gaps in
$\mbox{MgB}_2$
has been predicted theoretically by Liu et al.
\cite{LMK.01}
employing the effective $\sigma$ and $\pi$ two-band model.
They have obtained gaps at $T\!=\!0$,
$\Delta _{\sigma } \!=\! 7.4$ [meV]
and
$\Delta _{\pi } \!=\! 2.4$ [meV]
and also
their temperature dependencies and
$T_c\!=\!40$K.
The two-band model was first proposed by Suhl et al.
\cite{SMW.59}
and next introduced by Kondo
\cite{Kondo.63}.

In spite of theoretical great successes
by the two-band model and
the Eliashberg's strong-coupling theory,
the resonating mean-field theory (Res-MFT)
\cite{Fuku.88,NishiFuku.91}
may stand as a candidate for a possible theory and is considered to be useful
for such a theoretical approach.
Fermion systems with large quantum fluctuations
show serious difficulties in many-body problems at finite temperature.
To approach such problems,
Fukutome has developed the Res-Hartree-Fock theory (Res-HFT)
\cite{Fuku.88}
and Fukutome and one of the present authors (S.N.)
have extended it directly to
the Res-Hartree-Bogoliubov theory (Res-HBT)
to include pair correlations
\cite{NishiFuku.91,NishiFuku.92},
basing on the Lie algebra $U(N)$ and $SO(2N)$
of fermion pair operators (N: number of single-particle states),
respectively.
Steadily
the Res-HBT has succeeded to describe effectively the two-gap SC
\cite{NPO.06}
(referred to as I).
If we get a {\it Thermal Gap Equation} in the Res-HBT,
it is a strong manifestation of analogy of
the Res-HBT with the usual BCS and HBT
\cite{BCS.57,Bogo.59,deGennes.66,RingSchuk.80}.
The Res-HBT has a surprising fact that
every HB eigenfunction in a Res-HB state
has its own orbital-energy.
Due to this fact,
thus the Res-HBT, namely
the Res-MFT,  is considered to be
a possible candidate for approaching to such subjects.
This is because that the Res-HBT has the following characteristic feature:
The Res-HBT is equivalent to the coupled Res-HB eigenvalue equations and
the orbital concept is still surviving in the Res-HB approximation (Res-HBA)
though the orbitals are resonating.
This feature permits us to say that in some sense
the band picture has a correspondence to the orbital concept in the Res-HBA
though bands of different structures are resonating.
The structure of the coupled Res-HB eigenvalue equations
resembles considerably the structure of
the coupled quasiclassic Usadel equations
\cite{Usadel.70}
derived by Gurevich
for an anisotoropic two-band superconductor
\cite{Gurevich.03}.
The Res-HB ground state generated with HB wave functions (WFs)
which are
the coherent state representations (CS reps)
\cite{Perelomov.72},
is expected to almost exhausts the ground-state correlation energy
in all the correlation regimes.
The generator coordinate method (GCM) is also a powerful tool  for such the problem.
The modern GCM is widely used in nuclear and molecular physics
\cite{BGGY.11,OCSU.07}.

To demonstrate the advantage of the Res-HBT for
superconducting fermion systems with large quantum fluctuations
over the usual BCS and Eliashberg theories,
we already have applied it to a naive BCS Hamiltonian of singlet-pairing.
A state with large quantum fluctuations
is approximated by superposition of two HB WFs
which are non-orthogonal CS reps with different correlation structures.
We have optimized directly the Res-MF energy functionals
by variations of the Res-MF ground-state energy
with respect to the Res-MF parameters, i.e., {\it energy-gaps}.
The Res-MF ground and excited states
generated with the two HB WFs
explain most of the magnitudes of two energy-gaps in $\mbox{MgB}_2$.
Both the large {\em energy-gap} and the small one
have a significant physical meaning
because electron systems,
composed of condensed electron pairs,
have now strong correlations among the fermions
\cite{NPO.06}.

To go beyond the above mentioned {\em ab initio}
density functional computations and phenomenologies,
we develop a thermal Res-HBT which enables us 
to describe exactly a superconducting fermion system
with $N$ single-particle states.
A thermal Res-Fock-Bogoliubov (Res-FB) operator
plays a central and crucial role in the thermal Res-HBT
\cite{Fuku.88,NishiFuku.91}.
Using such an operator,
a temperature dependent variation should be made necessarily
along a way different from the usual thermal-BCS theory
\cite{KA.59,AGD.65,Abrikosov.88}.
Let us now prepare a Res-HB subspace spanned by Res-HB
ground and excited states.
We also introduce the projection operator $P$ to the Res-HB subspace.
A partition function in a CS rep of
the Lie algebra $SO(2N)$
$| g \rangle$
\cite{Perelomov.72}
is expressed as
as
$\mbox{Tr}(e^{-\beta H}) 
\!\!=\!\!
2^{N\!-\!1} \! \int \! \langle g |e^{-\beta H}| g \rangle dg~
( \beta \!=\! 1/k_B T )$ 
where Tr means trace and
the integration is the group integration on the Lie group $SO(2N)$.
Making use of the projection operator $P$,
the partition function in the Res-HB subspace is given as
$\mbox{Tr}(Pe^{-\beta H})$.
This kind of trace formula is calculated within the Res-HB subspace
by using the Laplace transform of 
$e^{-\beta H}$
and the projection operator method
\cite{Naka.58,Zwan.60,Mori.65,Fulde.93}
which leads us to an infinite matrix continued fraction (IMCF).
For the moment such a trace formula is assumed to be calculated appropriately.
A group action on an HB-Hamiltonian and -density matrix
at finite temperature are exactly defined.
The variation of the Res-HB free energy
is made parallel to the usual thermal BCS theory
\cite{BCS.57,Bogo.59,KA.59,AGD.65,Abrikosov.88},
which leads to
a thermal HB density matrix
$W_{\mbox{{\scriptsize Res}}}
^{\mbox{{\scriptsize thermal}}}$
expressed in terms of
the thermal Res-FB operator
${\cal F}_{\mbox{{\scriptsize Res}}}
^{\mbox{{\scriptsize thermal}}}$
as
$
W_{\mbox{{\scriptsize Res}}}
^{\mbox{{\scriptsize thermal}}}
\!=\!\!
\left[
1_{\!2N} \!+\! \exp 
\{ \! \beta {\cal F} _{\mbox{{\scriptsize Res}}}
^{\mbox{{\scriptsize thermal}}}
\right]^{\!-1}
$.
Then the Res-HB coupled eigenvalue equation is extended to
the thermal Res-HB coupled eigenvalue equation in a formal way
whose eigenvalue is obtained by diagonalization of
the thermal Res-FB operator.
For the sake of simplicity
here the whole Res-HB subspace is assumed to be
superposition of two HB WFs.
In this simplest case
we apply the present tentative of the thermal Res-HBT
to the naive BCS Hamiltonian of singlet-pairing and then
derive formulas for determining
$T_{\!c}$ and thermal behaviors of the gaps near $T \!\!=\!\! 0$ and $T_{\!c}$.
Particularly in the case of {\em equal} magnitude of two gaps but with different phases,
we find new formulas for  $T_{\!c}$
boosting up $T_{\!c}$ to higher values than the usual HB values 
and get new analytical expressions for gaps near $T \!\!=\!\! 0$ and $T_{\!c}$.
In the intermediate temperature region,
we solve a thermal resonating gap equation numerically
and show temperature dependence of the solved gaps.

To tackle the difficult problem mentioned above,
we are forced to propose a more rigorous thermal Res-HBA.
We hope that
based on projection-operator method
we can construct a strict thermal-Res-HBT and give
another MF approximation.
A calculation of the partition function by the IMCF, however, is difficult and
a procedure of tractable optimization is
too complicated to perform.
As a realistic problem,
it is better to seek for another possible and more practical way of
computing approximately the partition function and the Res-HB free energy
within the framework of the Res-MFT.
For this aim, it may be useful to introduce a quadratic Res-HB Hamiltonian
consisting of Res-FB operators.

In Section 2, 
the Res-HB coupled eigenvalue equation is extended to
the thermal Res-HB coupled eigenvalue equation in a formal way
and the expression for thermal HB density matrix is given
in terms of the thermal Res-FB operator.
In Section 3,
for simplicity the whole Res-HB subspace is assumed to be
superposition of two HB WFs.
Then 
we get a thermal resonating gap equation and
derive formulas for
$T_{\!c}$ and the gap near $T \!\!=\!\! 0$ and $T_{\!c}$.
For an {\em equal} magnitude of two gaps,
we find new formulas for  $T_{\!c}$.
In the intermediate temperature region,
we solve the thermal resonating gap equation numerically.
Finally in the last Section,
we give a summary and further perspectives.
In Appendices
we give a proof of trace formula and a derivation of
the expression for thermal HB density matrix
in terms of thermal Res-FB operator
We further provide the formulas
to calculate the gap at zero and intermediate temperatures.

\newpage

%%%%%%%%%%%%%%%%%%%%%%%%%%%%%
%                                                                                 %
%  2   Thermal resonating HB eigenvalue equation     %
%                                                                                 %
%%%%%%%%%%%%%%%%%%%%%%%%%%%%%

\def\thesection{\arabic{section}}
\setcounter{equation}{0}
\renewcommand{\theequation}{\arabic{section}.\arabic{equation}}

\section{Thermal resonating HB eigenvalue equation}

~~~~~~According to the principles of quantum statistical physics,
the free energy $F$ is given
in terms of the statistical density matrix 
$\stackrel{\circ }{W}$
as follows:%\\[-12pt]
\beq
F 
= 
\mbox{Tr}(\stackrel{\circ }{W}\! H) 
+ \frac{1}{\beta }
\mbox{Tr}(\stackrel{\circ }{W}\! \ln \stackrel{\circ }{W}),~~
\stackrel{\circ }{W} 
= 
\frac{e^{-\beta H}}{\mbox{Tr}(e^{-\beta H})} .
%\left( \! \beta \!=\! \displaystyle{\frac{1}{k_B T}} \! \right)  .
\label{exactfreeenergy}
\eeq\\[-20pt]

Consider the whole Res-HB subspace
$
|\Psi ^{\mbox{{\scriptsize Res}}(k)} \rangle
\!=\!
\sum _{r=1}^n
c_r ^{(k)}
|g _r \rangle~
(k \!=\! 1, \cdots , n)
$,
namely
the superposition of HB WFs
$|g _r \rangle$,
in which the Res-state with index
$k \!=\! 1$
and
the Res-states with indices
$k \!=\! 2, \!\cdots\!, n$
stand for 
the Res-ground state and the Res-excited states, respectively.
Let us introduce a projection operator $P$ to
the Res-HB subspace,
$P |\Psi \rangle = |\Psi ^{\mbox{{\scriptsize Res}}} \rangle$,
as%\\[-16pt]
\beqa
\left.
\begin{array}{cc}
P
\equiv
&\!\!\!
\sum _{r,s=1}^n |g _r \rangle (S^{-1})_{rs} \langle g _s |
=
P^\dagger, ~~Q = 1 - P ,\\
\\[-8pt]
&\!\!\!
P^2 = P ,~~Q^2 = Q ,~~PQ = QP = 0 ,
\end{array}
\right\}
\label{projectionoperator}
\eeqa\\[-8pt]
where 
$
S
\!=\!
(S_{rs})
\left(
\!=\! 
[ \det z_{rs}]^{\frac{1}{2}}
\right)
$
is an $n \!\times\! n$ matrix composed of the overlap integrals
and $S^\dagger \!=\! S$.
Here we propose a quantum statistical Res-HB theory
along the same way as the Peierls-Bogoliubov's quantum statistical  approach
\cite{Peierls.55,Bogo.59,BTS.59}.
Using the projection operator $P$,
we can extend
the HB free energy corresponding to the form of the free energy
(\ref{exactfreeenergy})
naturally  to
the Res-HB free energy in the following form:\\[-8pt]
\beq
F_{\mbox{{\scriptsize Res}}}
\!=\!
\mbox{Tr}(\stackrel{\circ }{W}_{\mbox{{\scriptsize Res}}} H)  
\!+\!
\frac{1}{\beta }
\mbox{Tr}
\left\{
\stackrel{\circ }{W}_{\mbox{{\scriptsize Res}}}
\ln
\stackrel{\circ }{W}_{\mbox{{\scriptsize Res}}} \!
\right\},~
\stackrel{\circ }{W}_{\mbox{{\scriptsize Res}}}
\equiv
\frac{Pe^{-\beta H}P}
{\mbox{Tr}(Pe^{-\beta H})} ,
\label{Res-HBfreeenergy}
\eeq\\[-8pt]
which leads directly to\\[-6pt]
\beq
F_{\mbox{{\scriptsize Res}}}
\!=\! 
\langle H \rangle_{\mbox{{\scriptsize Res}}}
\!+\!
\frac{1}{\beta }
{\displaystyle
\frac{
\mbox{Tr} \!
\left\{ \!
Pe^{- \beta H}\!P \ln (Pe^{- \beta H}\!P) \!
\right\}
}
{\mbox{Tr} (Pe^{- \beta H})}
}
- \frac{1}{\beta } \ln \mbox{Tr}
(Pe^{- \beta H}),~
\langle H \rangle_{\mbox{{\scriptsize Res}}}
\!\equiv\!
{\displaystyle
\frac{\mbox{Tr}(Pe^{- \beta H}\!PH)}
{\mbox{Tr} (Pe^{- \beta H})}
} .
\label{Res-HBfreeenergy2}
\eeq\\[-8pt]
In the denominator of resonating statistical density matrix 
$\stackrel{\circ }{W}_{\mbox{{\scriptsize Res}}}$
(\ref{Res-HBfreeenergy})
and in that of Res-HB free energy
$F_{\mbox{{\scriptsize Res}}}^{\mbox{\scriptsize thermalHB}}$
(\ref{Res-HBfreeenergy2}),
there appears
the partition function in the Res-HB subspace,
which is computed as\\[-20pt]
\beqa
\begin{array}{c}
\mbox{Tr} (Pe^{- \beta H})
=
\sum _{r,s=1}^n
\langle g _r |
e^{- \beta H}
|g _s \rangle (S^{-1})_{sr} ,
\end{array}
\label{traceformP}
\eeqa\\[-18pt]
the datailed proof of which is given in Appendix A.

On the other hand,
using the entropy
$S_{\mbox{{\scriptsize Res}}}^{\mbox{\scriptsize thermalHB}}$
in the Res-HB subspace
and the relation
$
F_{\mbox{{\scriptsize Res}}}^{\mbox{\scriptsize thermalHB}}
=
\langle H \rangle _{\mbox{{\scriptsize Res}}}
-
T S_{\mbox{{\scriptsize Res}}}^{\mbox{\scriptsize thermalHB}}
$,
we have another form of the Res-HB free energy,
i.e., a well-known formula
expressed in terms of a thermal Res-HB density matrix
$W_{\mbox{{\scriptsize Res}}}^{\mbox{{\scriptsize thermal}}}$
as\\[-14pt]
\beqa
\!\!\!\!
\left.
\begin{array}{c}
F_{\mbox{{\scriptsize Res}}}^{\mbox{\scriptsize thermalHB}}
\!=\!
\langle H \rangle _{\mbox{{\scriptsize Res}}}
\!+\!
{\displaystyle \frac{1}{2}\frac{1}{\beta }}
\mbox{Tr}
\left\{ \!
W_{\mbox{{\scriptsize Res}}}^{\mbox{{\scriptsize thermal}}}
\ln W_{\mbox{{\scriptsize Res}}}^{\mbox{{\scriptsize thermal}}}
\!+\!
(1_{2N} \!-\! W_{\mbox{{\scriptsize Res}}}^{\mbox{{\scriptsize thermal}}})
\ln (1_{2N} \!-\! W_{\mbox{{\scriptsize Res}}}^{\mbox{{\scriptsize thermal}}}) \!
\right\},\\
\\[-4pt]
W_{\mbox{{\scriptsize Res}}}^{\mbox{{\scriptsize thermal}}} 
 \!\equiv\!
\left[  \! \!
\begin{array}{cc}
R_{\mbox{{\scriptsize Res}}}^{\mbox{{\scriptsize thermal}}}
&K_{\mbox{{\scriptsize Res}}}^{\mbox{{\scriptsize thermal}}}\\
\\
-K_{\mbox{{\scriptsize Res}}}^{\mbox{{\scriptsize thermal}}\ast }&
1_N - R_{\mbox{{\scriptsize Res}}}^{\mbox{{\scriptsize thermal}}\ast}
\end{array}  \! \!
\right] ,
\begin{array}{c}
R_{\mbox{{\scriptsize Res}}}^{\mbox{{\scriptsize thermal}}}
 \!= \!
(R_{\mbox{{\scriptsize Res}};\alpha \beta }^{\mbox{{\scriptsize thermal}}}) ,
\\ \\
K_{\mbox{{\scriptsize Res}}}^{\mbox{{\scriptsize thermal}}}
 \!= \!
(K_{\mbox{{\scriptsize Res}};\alpha \beta }^{\mbox{{\scriptsize thermal}}}) .
\end{array}  \!\!\!
\end{array}  \!\!\!
\right\}
\label{densityform}
\eeqa\\[-8pt]
where,
using the resonating statistical density matrix
$\stackrel{\circ }{W}_{\mbox{{\scriptsize Res}}}$,
the quantities
$R_{\mbox{{\scriptsize Res}};\alpha \beta }^{\mbox{{\scriptsize thermal}}}$
and
$K_{\mbox{{\scriptsize Res}};\alpha \beta }^{\mbox{{\scriptsize thermal}}}$
are defined as\\[-8pt]
\begin{equation}
R_{\mbox{{\scriptsize Res}};\alpha \beta }^{\mbox{{\scriptsize thermal}}}
\!\equiv \!
\mbox{Tr}
\left\{ \!
\stackrel{\circ }{W}_{\mbox{{\scriptsize Res}}} \!
\left( \!
{E^{\beta } }_{\alpha }
+
\frac{1}{2}{\delta }_{\beta \alpha } \!
\right) \!
\right\}  ,~
K_{\mbox{{\scriptsize Res}};\alpha \beta }^{\mbox{{\scriptsize thermal}}}
 \!\equiv \!
\mbox{Tr}
\left\{
\stackrel{\circ }{W}_{\mbox{{\scriptsize Res}}} \!
E_{\beta \alpha }
\right\} .
\label{matrixelementsRK}
\end{equation}

Applying the trace manipulation,
the last equation of 
(\ref{traceform4}),
developed in Appendix A,
to the trace formulas
(\ref{matrixelementsRK}),
the second equation in
(\ref{Res-HBfreeenergy2})
can be expressed as\\[-18pt]
\beqa
\langle H \rangle
_{\mbox{{\scriptsize Res}}}
=
h_{\beta \alpha }
\mbox{Tr}
\left\{ \!
\stackrel{\circ}{W}_{\mbox{{\scriptsize Res}}} \!
\left( \!
E_{\;\,\alpha }^\beta
+
\frac{1}{2}
{\delta }_{\beta \alpha } \!
\right) \!
\right\}
+ 
{\displaystyle \frac{1}{4}}
[\alpha \beta |\gamma \delta]
\mbox{Tr}
\left\{
\stackrel{\circ}{W}_{\mbox{{\scriptsize Res}}} \!
E^{\alpha \gamma } E_{\delta \beta }
\right\} ,
\label{statisticalexpectationvalueofH1}
\eeqa\\[-20pt]
where
\beqa
\left.
\!\!\!\!\!\!
\begin{array}{rl}
&\mbox{Tr} \!
\left\{ \!
\stackrel{\circ }{W}_{\mbox{{\scriptsize Res}}} \!
\left( \!
E_{\;\,\alpha }^\beta
\!+\!
{\displaystyle \frac{1}{2}}
{\delta }_{\beta \alpha } \!
\right) \!
\right\}
\!=\!
\sum_{k=1}^n
\sum_{r,s=1}^n
{\displaystyle
\frac{c_r ^{(k)*} c_s^{(k)}}
{\mbox{Tr} (Pe^{- \beta H})}
}
\sum_{r^{\prime },s^{\prime } =1}^n
\langle g_r|
e^{- \beta H} \!
|g_{r^\prime } \rangle
\left( S^{-1} \right)_{r^{\prime } s^{\prime }} \\
&~~~~~~~~~~~~~~~~~~~~~~~~~~~~~~~~~~~~~~~~~~~
~~~~~~~~~~~~~~~~~~~~~~~~~~~~~~~~~
\times
\langle g_{s^{\prime }} |
{E^{\beta } }_{\alpha }
\!+\!
{\displaystyle \frac{1}{2}}
{\delta }_{\beta \alpha } \!
|g_s \rangle ,\\
\\[-12pt]
&\mbox{Tr} \!
\left\{ \!
\stackrel{\circ}{W}_{\mbox{{\scriptsize Res}}} \!
E_{\beta \alpha } \!
\right\}
\!=\!
\sum_{k=1}^n
\sum_{r,s=1}^n
{\displaystyle
\frac{c_r ^{(k)*} c_s^{(k)}}
{\mbox{Tr} (Pe^{- \beta H})}
}
\sum_{r^{\prime },s^{\prime } =1}^n
\langle g_r|
e^{- \beta H} \!
|g_{r^{\prime }} \rangle
\left( S^{-1} \right)_{r^{\prime } s^{\prime }}
\langle g_{s^{\prime }}|
E_{\beta \alpha }
|g_s \rangle .
\end{array}
\right\}
\label{statisticalexpectationvalueofH2}
\eeqa
The relation
$
\sum_{k=1}^n
c_r ^{(k)*} c_s^{(k)}
\!=\!
(S^{-1})_{sr}$
is satisfied
if the thermal Res-HB CI equation,
which is given later, 
could be solved and
all the mixing coefficients could be determined completely.
We have a simpler expression for
$\langle H \rangle
_{\mbox{{\scriptsize Res}}}$,
redenoted
as
$\langle H \rangle
_{\mbox{{\scriptsize Res}}}
^{\mbox{{\scriptsize thermal}}}$,
in the following form:\\[-12pt]
\beqa
\left.
\begin{array}{c}
\langle H \rangle
_{\mbox{{\scriptsize Res}}}
^{\mbox{{\scriptsize thermal}}}
=
\sum_{k=1}^n
\sum_{r,s=1}^n
H
\left[
W_{\mbox{{\scriptsize Res}}:rs}
^{\mbox{{\scriptsize thermal}}}
\right]
\cdot
\left[
\det z_{rs}
\right]^{\frac{1}{2}}
{\displaystyle
\frac{c_r ^{(k)*} c_s^{(k)}}
{\mbox{Tr} (Pe^{- \beta H})}
} ,\\
\\%[-4pt]
\begin{array}{rl}
W_{\mbox{{\scriptsize Res}}:rs}
^{\mbox{{\scriptsize thermal}}}
\equiv
\left[ \!\!
\begin{array}{cc}
R_{\mbox{{\scriptsize Res}}:rs}
^{\mbox{{\scriptsize thermal}}}&
K_{\mbox{{\scriptsize Res}}:rs}
^{\mbox{{\scriptsize thermal}}}\\
\\
-K_{\mbox{{\scriptsize Res}}:sr}
^{\mbox{{\scriptsize thermal}}\ast } &
1_N - R_{\mbox{{\scriptsize Res}}:sr}
^{\mbox{{\scriptsize thermal}}\ast }
\end{array} \!\!
\right] ,
\end{array}
\end{array}
\right\}
\label{statisticalexpectationvalueofH3}
\eeqa
in which
the explcit form of the Hamiltonian matrix element
is given as\\[-16pt]
\beqa
\begin{array}{c}
H
\left[
W_{\mbox{{\scriptsize Res}}:rs}
^{\mbox{{\scriptsize thermal}}}
\right]
=
h_{\alpha \beta }
R_{\mbox{{\scriptsize Res}}:rs;\beta \alpha }
^{\mbox{{\scriptsize thermal}}} \\
\\[-6pt]
+ 
{\displaystyle \frac{1}{2}}
[\alpha\beta |\gamma\delta]
\left\{
R_{\mbox{{\scriptsize Res}};rs;\beta \alpha }
^{\mbox{{\scriptsize thermal}}}
R_{\mbox{{\scriptsize Res}}:rs;\delta \gamma }
^{\mbox{{\scriptsize thermal}}}
- 
{\displaystyle \frac{1}{2}}
K_{\mbox{{\scriptsize Res}}:sr;\alpha \gamma }
^{\mbox{{\scriptsize thermal}}\ast }
K_{\mbox{{\scriptsize Res}}:rs;\delta \beta }
^{\mbox{{\scriptsize thermal}}}
\right\} ,
\end{array}
\label{statisticalexpectationvalueofH4}
\eeqa\\[-12pt]
where\\[-20pt]
\beqa
\left.
\begin{array}{ll}
&R_{\mbox{{\scriptsize Res}}:rs;\alpha \beta }
^{\mbox{{\scriptsize thermal}}}
\cdot
[\det z_{rs}]^{\frac{1}{2}}
=
\sum_{r^{\prime },s^{\prime } =1}^n
\langle g_r|
e^{- \beta H} \!
|g_{r^\prime } \rangle
\left( S^{-1} \right)_{r^{\prime } s^{\prime }}
\langle g_{s^{\prime }} |
{E^{\beta } }_{\alpha }
\!+\!
{\displaystyle \frac{1}{2}}
{\delta }_{\beta \alpha } \!
|g_s \rangle, \\
\\%[-6pt]
&K_{\mbox{{\scriptsize Res}}:rs;\alpha \beta }
^{\mbox{{\scriptsize thermal}}}
\cdot
[\det z_{rs}]^{\frac{1}{2}}
=
\sum_{r^{\prime },s^{\prime } =1}^n
\langle g_r|
e^{- \beta H} \!
|g_{r^{\prime }} \rangle
\left( S^{-1} \right)_{r^{\prime } s^{\prime }}
\langle g_{s^{\prime }}|
E_{\beta \alpha }
|g_s \rangle .
\end{array}
\right\}
\label{statisticalmatrixelements}
\eeqa

To determine 
$|g_r \rangle$'s and $c_r^{(k)}$'s
by the variational method,
we adopt a thermal Lagrangian with
Lagrange multiplier term $E^{(k)}$ to secure
normalization condition 
$
\langle
\Psi ^{\mbox{{\scriptsize Res}}(k)} |
\Psi  ^{\mbox{{\scriptsize Res}}(k)}
\rangle
\!=\!
1$,%\\[-18pt]
\beqa
\begin{array}{c}
L_{\mbox{{\scriptsize Res}}}
^{\mbox{{\scriptsize thermalHB}}}
=
\sum_{k=1}^n
\sum_{r,s=1}^n
\left\{
H\left[
W_{\mbox{{\scriptsize Res}}:rs}
^{\mbox{{\scriptsize thermal}}}
\right] - E^{(k)}
\right\}
\cdot
\left[
\det z^{\mbox{{\scriptsize thermal}}}_{rs}
\right]^{\frac{1}{2}}
c_r ^{(k)*} c_s^{(k)} .
\end{array}
\label{statisticalLagrangian}
\eeqa
The variation of 
(\ref{statisticalLagrangian})
is made in a quite parallel manner
to the previous ones
\cite{Fuku.88,NishiFuku.91}.
From the variation of
$L_{\mbox{{\scriptsize Res}}}
^{\mbox{{\scriptsize thermalHB}}}$
with respect to $c^{(k) \ast}_{r}$ for ground state
$(k \!=\! 1)$
and
any $k$th excited state,
we get the thermal Res-HB CI equation 
to determine thermal mixing coefficients $c^{(k)}_{s}$
\beqa
\begin{array}{c}
\sum_{s=1}^n 
\left\{ 
H\left[
W_{\mbox{{\scriptsize Res}}:rs}
^{\mbox{{\scriptsize thermal}}}
\right] - E^{(k)} 
\right\} 
\cdot [\det z^{\mbox{{\scriptsize thermal}}}_{rs}]^{\frac{1}{2}}c^{(k)}_{s} 
= 0 .~(k = 1, \cdots, n)
\end{array}
\label{statisticalResCIequation}
\eeqa
We also make the variation of the Hamiltonian matrix element 
$H[W_{\mbox{{\scriptsize Res}}:rs}
^{\mbox{{\scriptsize thermal}}}]$
as

\beqa
\!\!\!\!
\left.
\begin{array}{cc}
\delta H[W_{\mbox{{\scriptsize Res}}:rs}
^{\mbox{{\scriptsize thermal}}}]
\!=\!
{\displaystyle \frac{1}{2}} {\mbox{Tr}} \!
\left\{ \!
{\cal F}[W_{\mbox{{\scriptsize Res}}:rs}
^{\mbox{{\scriptsize thermal}}}]
\delta W_{\mbox{{\scriptsize Res}}:rs}
^{\mbox{{\scriptsize thermal}}} \!
\right\} , ~
{\cal F}[W_{\mbox{{\scriptsize Res}}:rs}
^{\mbox{{\scriptsize thermal}}}]
\!=\!
\left[ \!
\begin{array}{cc}
F_{\mbox{{\scriptsize Res}}:rs}
^{\mbox{{\scriptsize thermal}}}&\!\!
D_{\mbox{{\scriptsize Res}}:rs}
^{\mbox{{\scriptsize thermal}}} \\
\\
-D_{\mbox{{\scriptsize Res}}:sr}
^{\mbox{{\scriptsize thermal}}\ast }&\!\!
-F_{\mbox{{\scriptsize Res}}:sr}
^{\mbox{{\scriptsize thermal}}\ast }
\end{array} \!
\right] \! , \!\!\! \\
\\[-6pt]
F_{\mbox{{\scriptsize Res}}:rs;\alpha \beta }
^{\mbox{{\scriptsize thermal}}} 
\!\equiv\!
{\displaystyle 
\frac{\delta 
H[W_{\mbox{{\scriptsize Res}}:rs}
^{\mbox{{\scriptsize thermal}}}]}
{\delta 
R_{\mbox{{\scriptsize Res}}:rs;\beta \alpha }
^{\mbox{{\scriptsize thermal}}}}
\!=\! 
h_{\alpha \beta }
\!+\!
[\alpha \beta |\gamma \delta]
R_{\mbox{{\scriptsize Res}}:rs;\delta \gamma }
^{\mbox{{\scriptsize thermal}}}
} ,\\
\\[-6pt]
D_{\mbox{{\scriptsize Res}}:rs;\alpha \beta }
^{\mbox{{\scriptsize thermal}}}
\!\equiv\!
{\displaystyle 
\frac{\delta 
H[W_{\mbox{{\scriptsize Res}}:rs}
^{\mbox{{\scriptsize thermal}}}]}
{\delta K 
_{\mbox{{\scriptsize Res}}:sr;\alpha \beta }
^{\mbox{{\scriptsize thermal}}\ast }}
\!=\! 
-\frac{1}{2}[\alpha \gamma |\beta \delta]
K_{\mbox{{\scriptsize Res}}:rs;\delta \gamma }
^{\mbox{{\scriptsize thermal}}}
}.
\end{array}
\right\}
\label{statisticaldeltaH}
\eeqa
The variation of 
the thermal HB interstate density matrix 
$W_{\mbox{{\scriptsize Res}}:rs}
^{\mbox{{\scriptsize thermal}}}$
and the overlap integral
$[\det z^{\mbox{{\scriptsize thermal}}}_{rs}]^{\frac{1}{2}}$
are also given by
\beqa
\left.
\begin{array}{cc}
\delta W_{\mbox{{\scriptsize Res}}:rs}
^{\mbox{{\scriptsize thermal}}}
\!=\!
D_{rs}(1_{2N} \!-\! W_{\mbox{{\scriptsize Res}}:rs}
^{\mbox{{\scriptsize thermal}}})
\!+\!
(1_{2N} \!-\! W_{\mbox{{\scriptsize Res}}:rs}
^{\mbox{{\scriptsize thermal}}})
\tilde{D}_{rs} ,\\
\\[-6pt]
D_{rs}
\!\equiv\!
u_s z^{\mbox{{\scriptsize thermal}}-1}_{rs} \delta u^\dagger _r ,
{\quad}
\tilde{D}_{rs} 
\!\equiv\! 
\delta u_s z^{\mbox{{\scriptsize thermal}}-1}_{rs} u^\dagger _r , \\
\\[-6pt]
\delta [\det z^{\mbox{{\scriptsize thermal}}}_{rs}]^{\frac{1}{2}}
\!=\!
{\displaystyle \frac{1}{2}} 
{\hbox{Tr}}(D_{rs} \!+\! \tilde{D}_{rs}) 
\cdot 
[\det z^{\mbox{{\scriptsize thermal}}}_{rs}]^{\frac{1}{2}} .
\end{array}
\right\}
\label{statisticaldeltaWanddetz}
\eeqa
Following I and the Res-HFT
\cite{Fuku.88},
writing
$
L_{\mbox{{\scriptsize Res}}}
^{\mbox{{\scriptsize thermalHB}}}
\! = \!
\sum_{k = 1} ^n
\sum_{r,s=1}^n
{\cal L}_{\mbox{{\scriptsize Res}}:rs}
^{\mbox{{\scriptsize thermalHB}}(k)}
c_r ^{(k)*} c_s^{(k)}
$
and
$
{\cal L}_{\mbox{{\scriptsize Res}}:rs}
^{\mbox{{\scriptsize thermalHB}}(k)}
\! = \!
\{ H[W_{\mbox{{\scriptsize Res}}:rs}
^{\mbox{{\scriptsize thermal}}}]
\!-\!
E^{(k)} \} 
\!\cdot\!
[\det z^{\mbox{{\scriptsize thermal}}}_{rs}]^{\frac{1}{2}}
$,
then from the variation of
$L_{\mbox{{\scriptsize Res}}}
^{\mbox{{\scriptsize thermalHB}}}$,
namely
(\ref{statisticaldeltaH})
and
(\ref{statisticaldeltaWanddetz}),
we obtain the thermal Res-HB equation to determine
the MF WF $u_r$'s as
\beqa
\left.
\!\!\!\!
\begin{array}{cc}
\sum_{k =1}^n
\sum_{s =1}^n
{\cal K}_{\mbox{{\scriptsize Res}}:rs}
^{\mbox{{\scriptsize thermal}}(k)}
c_r ^{(k)*} c_s^{(k)}
= 0, \\
\\
{\cal K}_{\mbox{{\scriptsize Res}}:rs}
^{\mbox{{\scriptsize thermal}}(k)}
\!\!\equiv\!\!
\left\{ \!
(1_{2N} \!\!-\!\! W_{\mbox{{\scriptsize Res}}:rs}
^{\mbox{{\scriptsize thermal}}})
{\cal F} [W_{\mbox{{\scriptsize Res}}:rs}
^{\mbox{{\scriptsize thermal}}}]
\!\!+\!\! 
H[W_{\mbox{{\scriptsize Res}}:rs}
^{\mbox{{\scriptsize thermal}}}]
\!\!-\!\!
E^{(k)} \!
\right\}
\!\cdot\! 
W_{\mbox{{\scriptsize Res}}:rs}
^{\mbox{{\scriptsize thermal}}}
\!\cdot\! 
[\det z^{\mbox{{\scriptsize thermal}}}_{rs}]^{\frac{1}{2}} , \!\!
\end{array}
\right \}
\label{thermalRes-HBequation}
\eeqa
from which we can derive
the following thermal Res-HB coupled eigenvalue equations:
\beqa
\left.
\!\!\!\!\!\!\!\!\!\!\!\!
\begin{array}{rl}
&
[{\cal F}_{\mbox{{\scriptsize Res}}:r}
^{\mbox{{\scriptsize thermal}}}u_r]_i
\!=\!
\epsilon_{ri}^{\mbox{{\scriptsize thermal}}}u_{ri},~~
\epsilon_{ri}^{\mbox{{\scriptsize thermal}}}
\!\equiv\! 
\widetilde{\epsilon}_{ri}^{\mbox{{\scriptsize thermal}}}
\!-\!
\sum_{k =1}^n \!
\left\{ \!
H[W_{\mbox{{\scriptsize Res}}:rr}
^{\mbox{{\scriptsize thermal}}}] \!-\! E^{(k)} \!
\right\} \!
|c_r ^{(k)}|^2 , \\
\\
&
{\cal F}_{\mbox{{\scriptsize Res}}:r}
^{\mbox{{\scriptsize thermal}}}
\!\equiv\! 
{\cal F}[W_{\mbox{{\scriptsize Res}}:rr}
^{\mbox{{\scriptsize thermal}}}] \!
\sum_{k =1}^n \! |c_r ^{(k)} |^2
\!+\!\!
\sum_{k =1}^n \!
\sum_{s=1}^{\prime~n} \!
\left\{ \!
{\cal K}_{\mbox{{\scriptsize Res}}:rs}
^{\mbox{{\scriptsize thermal}}(k)} \!
c_r ^{(k)*} c_s^{(k)} \!
\!+\! 
{\cal K}_{\mbox{{\scriptsize Res}}:rs}
^{\mbox{{\scriptsize thermal}}(k)\dag } \!
c_r ^{(k)} c_s^{(k)*} \!
\right\} . \!\!\!
\end{array}
\right\}
\label{thermalRes-HBeigenvalueequation}
\eeqa
We call the hermitian $2N \!\times\! 2N$ matrix
${\cal F}_{\mbox{{\scriptsize Res}}:r}
^{\mbox{{\scriptsize thermal}}}$
the thermal Res-FB operator.
Finally,
we derive the expression for thermal HB density matrix
in terms of thermal Res-FB operator.
From the relations
$[{\cal F}_{\mbox{{\scriptsize Res}}:r}
^{\mbox{{\scriptsize thermal}}},
W_{\mbox{{\scriptsize Res}}:rr}
^{\mbox{{\scriptsize thermal}}}] = 0$
(\ref{modificationofWFW2})
and
(\ref{equilibriumcondition}),
we can reach the $r$th thermal HB density matrix
$W_{\mbox{{\scriptsize Res}}:rr}
^{\mbox{{\scriptsize thermal}}}$,
which is expressed in terms of
the $r$th thermal Res-FB operator
${\cal F}_{\mbox{{\scriptsize Res}}:r}
^{\mbox{{\scriptsize thermal}}}$,
as
\beqa
W_{\mbox{{\scriptsize Res}}:rr}
^{\mbox{{\scriptsize thermal}}}
= 
\frac{1}
{1_{2N} + \exp 
\{ \beta {\cal F} _{\mbox{{\scriptsize Res}}:r}
^{\mbox{{\scriptsize thermal}}}
\}
} ,
\label{solutionWrr}
\eeqa
which is the generalizations of the usual thermal density matrix
\cite{Ozaki.85}
to the Res-MFT.
In this section, a thermal Res-HB theory has been developed
in a formal way.
%The more practical way to the thermal Res-HB theory is given in the succeeding Section.

For the moment the trace formula,
$\mbox{Tr}(Pe^{-\beta H})$
(\ref{traceformP})
is assumed to be computed suitably.
For our sake of simplicity,
in the next Section,
the whole Res-HB subspace is assumed to be
superposition of two HB WFs.
In this simplest case,
keeping an intimate connection with the usual BCS theory,
we apply a tentative of the thermal Res-HBT
to the naive BCS Hamiltonian of singlet-pairing and
derive formulas for determining
$T_{\!c}$ and thermal behaviors of the gaps near $T \!\!=\!\! 0$ and $T_{\!c}$.
We denote
$W_{\mbox{{\scriptsize Res}}:rs}^{\mbox{{\scriptsize thermal}}},~
{\cal F}_{\mbox{{\scriptsize Res}}:r}^{\mbox{{\scriptsize thermal}}}$
and
${\cal F} [W_{\mbox{{\scriptsize Res}}:rs}^{\mbox{{\scriptsize thermal}}}]$
simply as
$W_{rs},~{\cal F}_r$
and
${\cal F} [W_{rs}]$,
respectively.

\newpage

%%%%%%%%%%%%%%%%%%%%%%%
%                                                               %
%  3   Thermal resonating gap equation    %
%                                                               %
%%%%%%%%%%%%%%%%%%%%%%%

\def\thesection{\arabic{section}}
\setcounter{equation}{0}
\renewcommand{\theequation}{\arabic{section}.\arabic{equation}}

\section{Thermal resonating gap equation}

~~From
(\ref{solutionWrr})
the thermal HB density matrix is given as
$
W_{rrp}[{\cal F}_{rp}]
\! = \!
\left[
1 \!+\! \exp \{ \beta {\cal F}_{rp} \}
\right]^{-1} \!
(r\!=\!1,2)
$
in momentum $p$.
Using a Bogoliubov transformation $g_{1(2)p}$,
$W_{11(22)p}[{\cal F}_{1(2)p} ]$
are diagonalized as\\[-18pt]
\beqa
\widetilde{W}_{rp}
\!=\!
g_{rp} ^{\dagger } W_{rrp}[{\cal F}_{rp} ] g_{rp}
\!=\!
\left[ \!\!
\begin{array}{cc}
\widetilde{w}_{rp} &\!\! 0\\
\\[-10pt]
0 &\!\! 1 - \widetilde{w}_{rp}
\end{array} \!\!
\right] ,~
\widetilde{w}_{rp}
\!=\!
{\displaystyle
\frac{1}{1 \!+\!
e^{{\displaystyle \beta \widetilde{\epsilon }_{rp}}}}
},~
1 \!-\! \widetilde{w}_{rp}
\!=\!
{\displaystyle
\frac{1}{1 \!+\!
e^{{\displaystyle -\beta \widetilde{\epsilon }_{rp}}}}
} .
\label{tildeCalFrmatrix0}
\eeqa\\[-12pt]
By making the Bogoliubov transformation $g_{rp}$,
eigenvalues
$\widetilde{\epsilon }_{rp}$
are obtained by diagonalization of
the thermal Res-FB operators ${\cal F}_{rp}$
with additional terms
$( H[W_{rrp}] \!-\! E ) |c_{rp}|^2~(r\!=\!1,2)$.
The thermal HB interstate density matrix
in the whole Res-HB subspace
is given as the direct sum:\\[-18pt]
\beqa
\begin{array}{l}
W_p [{\cal F}_p]
\!=\!
g_p \widetilde{W_p}g_p ^{\dagger }
\!=\!
\bigoplus_{r=1}^2 W_{rrp}[{\cal F}_{rp}],~~
W_{rr}[{\cal F}_r ]
\!=\!
g_{rp} \widetilde{W}_{rp} g_{rp} ^{\dagger } .
\end{array}
\label{WrrCalFrmatrix}
\eeqa\\[-30pt]

Suppose a {\em tilde} thermal Res-HB density operator
$\widetilde{W}_{1(2)p}$
for {\em equal-gaps} to be\\[-18pt]
\beqa
\widetilde{W}_{1(2)p}
\!=\!
\left[ \!\!
\begin{array}{cc}
\widetilde{W}_{1(2)p} ^{\uparrow } \!\cdot\! I_2 & 0 \\
\\[-10pt]
0 & \widetilde{W}_{1(2)p} ^{\downarrow } \!\cdot\! I_2
\end{array} \!\!
\right] ,~~
\widetilde{W}_{1(2)p} ^{\uparrow (\downarrow)}
\!=\!
\left[ \!\!
\begin{array}{cc}
\widetilde{w}_{1(2)p} ^{\uparrow (\downarrow)} \!\cdot\! I_2 & 0 \\
\\[-10pt]
0 & (1 \!-\! \widetilde{w}_{1(2)p} ^{\uparrow (\downarrow)})
\!\cdot\! I_2
\end{array} \!\!
\right] .
\\[-12pt] \nonumber
\eeqa
Here $I_2$ is the two-dimensional unit matrix.
Performing the unitary transformation by
$\widehat{g}_{1(2)p} ^{\uparrow (\downarrow)}$,
we obtain the following thermal Res-HB density matrix
$\widetilde{W}_{1(2)p} ^{\uparrow (\downarrow)}$.\\[-18pt]
\beqa
\begin{array}{rl}
&\!\!\!\!\!\!\!\!
W_{1(2)p}^{\uparrow (\downarrow)}
\!=\!
\widehat{g}_{1(2)p} ^{\uparrow (\downarrow)}
\widetilde{W}_{1(2)p} ^{\uparrow (\downarrow)}
\widehat{g}_{1(2)p} ^{\uparrow (\downarrow)\dagger }
\!=\!
\widehat{g}_{1(2)p} ^{\uparrow (\downarrow)} \!
\left[ \!\!
\begin{array}{cc}
\widetilde{w}_{1(2)p} ^{\uparrow (\downarrow)} \!\cdot\! I_2 & 0 \\
\\[-10pt]
0 & (1 \!-\! \widetilde{w}_{1(2)p} ^{\uparrow (\downarrow)})
\! \!\cdot\! I_2 \!
\end{array} \!\!
\right] \!
\widehat{g}_{1(2)p} ^{\uparrow (\downarrow)\dagger } \\
\\[-6pt]
&\!\!\!\!\!\!\!\!
\!=\!\!
\left[ \!\!\!
\begin{array}{cc}
\frac{1}{2} \!
\left\{ \!
1 \!\!-\!\!
\cos \widehat{\theta } _{1(2)p}
\left( \!
1 \!\!-\!\! 2 \widetilde{w}_{1(2)p} ^{\uparrow (\downarrow)} \!
\right) \!
\right\}
\! \!\cdot\! I_2 \!
&\!\!\!\!\!\!
\{ \!-(+)\! \}  \! \times \!
\frac{1}{2} \!
\sin \widehat{\theta } _{1(2)p}e ^{-i\widehat{\psi } _{1(2)}} \!
\left( \!
1 \!\!-\!\! 2 \widetilde{w}_{1(2)p} ^{\uparrow (\downarrow)}
\right)
\! \!\cdot\! I_2 \!
\\[-6pt]
&\\
\{ \!-(+)\! \} \! \times \!
\frac{1}{2} \!
\sin \widehat{\theta } _{1(2)p}e ^{i\widehat{\psi } _{1(2)}} \!
\left( \!
1 \!\!-\!\! 2 \widetilde{w}_{1(2)p} ^{\uparrow (\downarrow)}
\right)
\! \!\cdot\! I_2 \!
&\!\!\!\!\!\!
\frac{1}{2} \!
\left\{ \!
1 \!\!+\!\!
\cos \widehat{\theta } _{1(2)p} \!
\left( \!
1 \!\!-\!\! 2 \widetilde{w}_{1(2)p} ^{\uparrow (\downarrow)} \!
\right) \!
\right\}
\! \!\cdot\! I_2 \!
\\
\end{array}
\!\!
\right] \! .
\end{array}
\label{Wrrmatrix2}
\eeqa\\[-10pt]
In the {\em equal-gaps} case
$(H[W_{11}] \!=\! H[W_{22}] \!=\! H[W])$,
following I,
the Res-HB ground (excited) energy
$E_{\mbox{\scriptsize gr}(\mbox{\scriptsize ex})}^{\mbox{\scriptsize Res}}$  
is classified into two cases,
according to the solutions for the Res-HB CI equation:\\[-12pt]
\beqa
\left.
\begin{array}{rcl}
\mbox{Case I}~~~&:&~~~H[W] \!-\! H[W_{12}]>0, \\
E_{\mbox{\scriptsize gr}(\mbox{\scriptsize ex})}^{\mbox{\scriptsize Res}}
&\!\!\!\! = \!\!\!\!&
{\displaystyle \frac{1}{1 \!+(-) [\det z_{12}]^{\frac{1}{2}}}}
\!\cdot\!
\left( \!
H[W] \!+(-) H[W_{12}] \!\cdot\!
[\det z_{12}]^{\frac{1}{2}} \!
\right), 
\end{array}
\right\}
\label{SolI}
\eeqa\\[-26pt]
\beqa
\left.
\begin{array}{rcl}
\mbox{Case II}~~~&:&~~~H[W] \!-\! H[W_{12}]<0, \\
E_{\mbox{\scriptsize gr}(\mbox{\scriptsize ex})}^{\mbox{\scriptsize Res}}
&\!\!\!\!\! = \!\!\!\!\!&
{\displaystyle \frac{1}{1 \!-(+) [\det z_{12}]^{\frac{1}{2}}}}
\!\cdot\!
\left( \!
H[W] \!-(+) H[W_{12}] \!\cdot\!
[\det z_{12}]^{\frac{1}{2}} \!
\right).
\end{array} \!
\right\}
\label{SolII}
\eeqa\\[-14pt]
From now we keep a close connection with the BCS theory,
especially in relation to the gap.
The Res-FB operator
${\cal F}^{\uparrow }_{1(2)p}$
for spin-up state,
accompanying quantities with upper or lower sign
corresponding to Case I
(\ref{SolI})
and Case II
(\ref{SolII}),
is expressed as\\[-18pt]
\beqa
{\cal F}^{\uparrow }_{1(2)p}
\!=\!
\left[ \!\!
\begin{array}{cc}
{\cal F}^{\uparrow }_{+ \varepsilon_p} \!\cdot\! I_2
&\{+ (-)\} \!\times\! {\cal F}^{\uparrow }_{\Delta p} \!\cdot\! I_2 \\
\\[-8pt]
\{+ (-)\} \!\times\! {\cal F}^{\uparrow }_{\Delta p} \!\cdot\! I_2
&-{\cal F}^{\uparrow }_{- \varepsilon_p} \!\cdot\! I_2
\end{array} \!\!
\right] .
\label{ResFB1p}
\eeqa\\[-12pt]
\vspace{-0.1cm}
The quantities
${\cal F}^{\uparrow }_{\!\!+ \varepsilon _p}$,$\!$
${\cal F}^{\uparrow }_{\!\!- \varepsilon _p}$
$\!$and$\!$
${\cal F}^{\uparrow }_{\!\!\Delta p}$
for Case I(upper sign) and Case II(lower sign) 
are defined as
\beqa
\!\!\!\!
\begin{array}{c}
{\cal F}^{\uparrow }_{\!\!+(-) \varepsilon _p}
\!\equiv\!
{\displaystyle \frac{1}{2}} \!
\left\{ \!
\varepsilon_p
\!+\!
2
\widetilde{E}_{\mbox{\scriptsize gr}}^{\mbox{\scriptsize Res}}
{\displaystyle 
\frac
{
\sin ^{2} \!
\frac{\theta _p}{2}
\!
\left( \!
\cos ^{2} \!
\frac{\theta _p}{2} \!
\right)
}
{\cos {\theta }_p}
\mp
\frac{\Delta ^2}{\varepsilon_p}}
\!\cdot\! [\det \! z_{12}]^{\frac{1}{2}} \!
\right\}
\!\cdot\!
{\displaystyle \frac{1}
{1 \!\!\pm\!\! [\det \! z_{12}]^{\frac{1}{2}}}} ,
\widetilde{E}_{\mbox{\scriptsize gr}}^{\mbox{\scriptsize Res}}
\!\equiv\!
H[W] \!-\!
E_{\mbox{\scriptsize gr}}^{\mbox{\scriptsize Res}},
\end{array}
\label{ResFepandFdp1}
\eeqa
\vspace{-0.8cm}
\beqa
\begin{array}{c}
{\cal F}^{\uparrow }_{\!\Delta p}
\!=\!
{\cal F}^{\uparrow }_{\!\Delta }
\!\equiv\!
-{\displaystyle
\frac{1}{2}} \!
\Delta \!
\left\{ \!
N(0) V
\!\cdot\! \mbox{arcsinh}
\left( \!
{\displaystyle \frac{1}{x}} \!
\right)
\pm
[\det \! z_{12}]^{\frac{1}{2}} \!
\right\}
\!\cdot\!
{\displaystyle \frac{1}
{1 \!\!\pm\!\! [\det \! z_{12}]^{\frac{1}{2}}}},
\left(\!
x
\!=\!
{\displaystyle \frac{\Delta }{\hbar \omega_D}} \!
\right) ,
\end{array}
\label{ResFepandFdp2}
\eeqa\\[-18pt]
where\\[-30pt]
\beqa
\!\!\!\!\!\!\!\!\!\!
\begin{array}{c}
[\det{z}_{12}]^{\frac{1}{2}} 
\!=\!
\exp
\left[
- 2 N(0) \hbar \omega_D
\left\{
{\displaystyle
\ln (1 \!+\! x^2 ) \!+\! 2 x
\!\cdot\! \arctan \!
\left( \!
\frac{1}{x} \!
\right)
} \!
\right\}
\right] .
\end{array} \!\!\!
\label{detz}
\eeqa\\[-18pt]
At finite temperature,
the quantities
${\cal F}^{\uparrow }_{\!+ \varepsilon _p}$,
${\cal F}^{\uparrow }_{\!- \varepsilon _p}$,
${\cal F}^{\uparrow }_{\!\Delta p}$
and
$\Delta$
become temperature-dependent.
This is explicitly expressed by a subscript $T$.
Using the distributions
(\ref{tildeCalFrmatrix0})
we require correspondence relations
$\cos \theta _{pT}
\!\Rightarrow\!
\cos \widehat{\theta } _{1p}$
and
$\sin \theta _{pT}
\!\Rightarrow\!
\sin \widehat{\theta } _{1p}$
given through\\[-18pt]
\beqa
\!\!\!\!
\left.
\begin{array}{c}
{\displaystyle
\cos \theta _{p}
\!=\!
\frac{\varepsilon _p}
{\sqrt{\varepsilon _p ^2 \!+\! \Delta ^2 _{\!T}}}
\!=\!
\frac{
{\cal F}^{\uparrow }_{\! + \varepsilon_p T}
+
{\cal F}^{\uparrow }_{\! - \varepsilon_p T} \!
}
{2 ~\! \widetilde{\epsilon }_{1p} ^{\uparrow }} \!
\left( \!
1 - 2 \widetilde{w}_{1p} ^{\uparrow } \!
\right)
} , \\
\\[-18pt]
{\displaystyle
\sin \theta _{p}
\!=\!
\frac{\Delta _{\!T}}
{\sqrt{\varepsilon _p ^2 \!+\! \Delta ^2 _{\!T}}}
\!=\!
- \frac{{\cal F}_{\!\! \Delta _T} ^{\uparrow }}
{\widetilde{\epsilon }_{1p} ^{\uparrow }} \!
\left( \!
1 \!-\! 2 \widetilde{w}_{1p} ^{\uparrow } \!
\right) ,
}
\end{array}
\right\}
\!\rightarrow\!
\frac{\Delta _{\!T}}{\varepsilon _p }
\!=\!
-
\frac{{\cal F}_{\! \Delta _T} ^{\uparrow } \!\!
\left( \!
1 \!-\! 2 \widetilde{w}_{1p} ^{\uparrow } \!
\right)
}
{\displaystyle
{\frac{{\cal F}^{\uparrow }_{\! + \varepsilon_p T}
+
{\cal F}^{\uparrow }_{\!\! - \varepsilon_p T}}{2}
} \!
\left( \!
1 \!-\! 2 \widetilde{w}_{1p} ^{\uparrow } \!
\right)
} .
\label{equivalenceforcosandsin}
\eeqa\\[-12pt]
Notice the multiplication factor
$
1 \!\!-\!\! 2 \widetilde{w}_{1p} ^{\uparrow (\downarrow)}
$.
The
$
\widetilde{\epsilon }^{\uparrow }_{1p}
(
\!=\!
\widetilde{\epsilon }^{\uparrow }_p
)
$
is the quasi-particle (QP) energy:
$
\widetilde{\epsilon }^{\uparrow }_p
\!\!=\!\!
\{
( \!
{\cal F}^{\uparrow }_{\! + \varepsilon_p T}
\!+\!
{\cal F}^{\uparrow }_{\! - \varepsilon_p T} \!
)^2 \!\!
/ 4
\!+\!
{\cal F}_{\! \Delta _T} ^{\uparrow 2} \!
\}^{1/2}
$.
The first two equations in parenthesis of
(\ref{equivalenceforcosandsin})
are unified into the second single equation in
(\ref{equivalenceforcosandsin}).
From now
it is shown that equation
(\ref{equivalenceforcosandsin})
plays the role of the self-consistency condition
at finite temperature.
Dividing numerator and denominator, respectively
by
$(\varepsilon _{\!p} ^2 \!+\! \Delta ^2 _{\!T}) ^{3/2}$
in R.H.S. of the second equation in
(\ref{equivalenceforcosandsin}),
we have\\[-20pt]
\beqa
1
\!=\!
\frac{
{\displaystyle
\frac{\varepsilon _p ^2}
{(\varepsilon _p ^2 \!+\! \Delta ^2 _T) ^{\frac{3}{2}}} \!
\left( \!
- \frac{2{\cal F}_{\Delta _T} ^{\uparrow }}{\Delta _T} \!
\right)
} \!
\left( \!
1 \!-\! 2 \widetilde{w}_{p} ^{\uparrow }
\right)
}
{
{\displaystyle
\frac{\varepsilon _p}
{(\varepsilon _p ^2 + \Delta ^2 _T) ^{\frac{3}{2}}}
} \!
\left( \!
{\cal F}^{\uparrow }_{+ \varepsilon_p T}
+
{\cal F}^{\uparrow }_{- \varepsilon_p T} \!
\right) \!
\left( \!
1 \!-\! 2 \widetilde{w}_{p} ^{\uparrow } \!
\right)
} ,~~~
\widetilde{w}_{p} ^{\uparrow }
\!=\!
\widetilde{w}_{1p} ^{\uparrow } .
\label{equivalenceforsin3}
\eeqa\\[-14pt]
Now we demand a new condition for
{\it Thermal Gap Equation}\\[-20pt]
\beqa
\begin{array}{c}
\sum_p \!
\left\{ \!
{\displaystyle
\frac{\varepsilon _p}
{
(\varepsilon _p ^2 \!+\! \Delta ^2 _T) ^{\frac{3}{2}}
}
{\displaystyle \frac{1}{2}} \!
\left( \!
{\cal F}^{\uparrow }_{+ \varepsilon_p T}
+
{\cal F}^{\uparrow }_{ - \varepsilon_p T} \!
\right)
}
-
{\displaystyle
\frac{\varepsilon _p ^2}
{
(\varepsilon _p ^2 \!+\! \Delta ^2 _T) ^{\frac{3}{2}}
} \!
\left( \!
- \frac{{\cal F}_{\Delta _T} ^{\uparrow }}{\Delta _T} \!
\right)
} \!
\right\} \!
\left(
1 \!-\! 2 \widetilde{w}_{p} ^{\uparrow }
\right)
\!=\! 0 ,
\end{array}
\label{gapequation}
\eeqa\\[-12pt]
which leads to\\[-20pt]
\beqa
\left.
\begin{array}{ll}
&\left\{ \!
1 \!-\! N(0) V
\!\cdot\! \mbox{arcsinh} \!
\left( \!
{\displaystyle \frac{1}{x_T}} \!
\right)
\!\mp\!
[\det z_{12}]_T ^{\frac{1}{2}} \!
\right\} \!
\sum _p \! A_p  \\
\\[-14pt]
&~~~~~~~~~~~~~~~~~~~~~~~~~~~~~~~~\!
+
\widetilde{E}_{\mbox{\scriptsize gr}_T}^{\mbox{\scriptsize Res}(\pm)}
\hbar \omega_D \!
\sum _p \! B_p
\!\mp\!
\Delta ^2 _T \!\cdot\! [\det z_{12}]_T ^{\frac{1}{2}}
\sum _p \! C_p
\!=\! 0 , \\
\\[-14pt]
&{\displaystyle
\frac{
\widetilde{E}_{\mbox{\scriptsize gr}_T}^{\mbox{\scriptsize Res}(\!\pm\!)}}
{\hbar \omega_D}
\!=\!
\pm 2 \!
N \! (0) \hbar \omega_D x_{\!T} ^2
\!\cdot\! \mbox{arcsinh} \!
\left( \!
\frac{1}{x_{\!T}} \!
\right) \!
\left\{ \!
2 \! - \! N \! (0) \! V \!
\!\cdot\! \mbox{arcsinh} \!
\left( \!
\frac{1}{x_{\!T}} \!
\right) \!\!
\right\} \!
\frac{[\det z_{12}]_T ^{\frac{1}{2}}}
{1
\!\pm\!
[\det z_{12}]_T ^{\frac{1}{2}}} \!
} ,
\end{array}
\right\}
\label{TempSCFcondition}
\eeqa\\[-14pt]
which we have calculated
using the solutions for the Res-HB CI equation obtained in I.
We also give the following definitions for
$\sum _p \! A_p,~\sum _p \! B_p$ and $\sum _p \! C_p$:\\[-18pt]
\beqa
\begin{array}{c}
\left[ \!
\sum _p \! A_p,~\sum _p \! B_p,~\sum _p \! C_p \!
\right]
\equiv
\sum _p \!
{\displaystyle
\left[ \!
\frac
{\varepsilon _p ^2}
{(\varepsilon _p ^2 \!+\! \Delta ^2 _T) ^{\frac{3}{2}}},~
\frac
{1}
{\varepsilon _p ^2 \!+\! \Delta ^2 _T},~
\frac
{1}
{(\varepsilon _p ^2 \!+\! \Delta ^2 _T) ^{\frac{3}{2}}} \!
\right] \!
\left( \!
1 \! - \! 2 \widetilde{w}_{p} ^{\uparrow } \!
\right)
}.
\end{array}
\label{DefinitionsofABandC}
\eeqa\\[-14pt]
Rearranging
(\ref{TempSCFcondition}),
it is cast to a {\it Thermal Gap Equation} with similar form to the one in I:\\[-20pt]
\beqa
{\displaystyle
\frac{1}{N(0) V}
}
\!=\!&\!\!\!\!\!\!\!\!\!\!\!\!\!\!\!\!\!\!\!\!\!\!\!\!\!\!\!\!\!
{\displaystyle
\mbox{arcsinh} \!
\left( \!
\frac{1}{x_T} \!
\right) \!\!
\left[
1 \!\pm\!
2 N(0) \!
\frac{\mbox{arcsinh} \!
\left( \!
\frac{1}{x_T} \!
\right)
}
{\sum _p \! A_p}
\frac{\Delta _T \!\cdot\! \Delta _T \!
\sum _p \! B_p
}
{1 \!\pm\! [\det z_{12}]_T ^{\frac{1}{2}}}
\!\cdot\!
[\det z_{12}]_T ^{\frac{1}{2}}
\right]
}
\nonumber \\
\nonumber \\[-10pt]
&\!\!\!\!\!
\!\times\!\!
\left[
1
\!+\!
\left\{ \!
{\displaystyle
\mp 1
\mp
\frac{\Delta ^2 _T \! \sum _p \! C_p}
{\sum _p \! A_p}
\!\pm\! 4 \! N(0) \!
\frac{\mbox{arcsinh} \!
\left( \!
\frac{1}{x_T} \!
\right)
}
{\sum _p \! A_p}
\frac{\Delta_{\!T} \!\cdot\! \Delta _T \!
\sum _p \! B_p
}
{1 \!\pm\! [\det z_{12}]_T ^{\frac{1}{2}}}
} \!
\right\}
\! \cdot \!
[\det z_{12}]_T ^{\frac{1}{2}}
\right] ^{\!-1}\!\!.
\label{GapequationwithABandC}
\eeqa\\[-12pt]
The summations
$\sum _p \! A_p,\sum _p \! B_p\!$ and $\!\sum _p \! C_p$ near $T \!=\! 0$
are computed in Appendix A.
Substituting
(\ref{TdependenceofABandC}), (\ref{Approxmations}) and (\ref{Approxmations2})
into (\ref{GapequationwithABandC}),
then, near $T \!=\! 0$ we have the gaps for Case I
(\ref{SolI}) as
\newpage
\beqa
{\displaystyle
\Delta _T ^{\mbox{\scriptsize I}}
=
\Delta_0 \!
\left[
1
-
\frac{[\det z_{12}]^{\frac{1}{2}}_0}{N(0) V}
\left\{
{\displaystyle
\mbox{arcsinh}
\left( \!
\frac{\hbar \omega_D}{\Delta_0} \!
\right)
- 1 +
\frac{[\det z_{12}]^{\frac{1}{2}}_0}{N(0) V}
}
\right\} ^{-1}
\cdot
T  ^{\mbox{\scriptsize I}}_{\frac{1}{2}}
\right]
} \! .
\label{DeltapnearT0}
\eeqa\\[-10pt]
On the other hand,
with the aid of
$
[\det z_{12}]^{1 / 2}_0
\!\simeq\!
1 \!-\! 2 \pi \! N(0) \hbar \omega_D x_0
(0 \!<\! x_0 \!\ll\! 1)
$
which is easily derived from Taylor expansion of
(\ref{detz}),
we obtain the gap for
Case II
(\ref{SolII})
as\\[-18pt]
\beqa
{\displaystyle
\!\!\!\!
\Delta _T ^{\mbox{\scriptsize II}}
=
\Delta_0 \!
\left[
1
\!+\!
\frac{1}{\pi \sqrt{1 \! + \!   (N(0) V) ^2}} \!
\left\{
\pi
\! - \!
\left(
1
\!\! - \!\!
N(0) V
\!\! - \!\!
\sqrt{1 \! + \!   (N(0) V) ^2}
\right) \!
\mbox{arcsinh}  \!
\left( \!
\frac{\hbar \omega_D}{\Delta_0} \!
\right) \!
\right\} \!
\cdot
T  ^{\mbox{\scriptsize II}}_{\frac{1}{2}}
\right] \! ,
}
\label{DeltamnearT0}
\eeqa\\[-10pt]
where we have used the approximate relation\\[-16pt]
\beqa
{\displaystyle
\frac{1}{2}
N(0) V
\mbox{arcsinh}  \!
\left( \!
\frac{\hbar \omega_D}{\Delta_0} \!
\right) \!\!
}
\left\{ \!
\mbox{arcsinh}  \!
\left( \!
{\displaystyle \frac{\hbar \omega_D}{\Delta_0} } \!
\right)
\!\! - \!\!
1
\! - \!
1
\!
\right\}
\approx 
\mbox{arcsinh}  \!
\left( \!
{\displaystyle \frac{\hbar \omega_D}{\Delta_0} } \!
\right)
\!\! - \!\!
1 .
\label{approrela}
\eeqa\\[-12pt]
In
(\ref{DeltapnearT0}) and (\ref{DeltamnearT0}),
as shown from Appendix A,
the quantity
$
T  ^{
{\mbox{\scriptsize I}}({\mbox{\scriptsize II}})
}_{\frac{n}{2}}
$
is defined by\\[-18pt]
\beqa
{\displaystyle
T  ^{
{\mbox{\scriptsize I}}({\mbox{\scriptsize II}})
}_{\frac{n}{2}}
\!\equiv\!
\sqrt{2 \pi}
\left\{ \!
\widetilde{\Delta}_T ^{
{\mbox{\scriptsize I}}({\mbox{\scriptsize II}})
} \!
\right\} ^{\! -\frac{n}{2}} \!
\left\{ \!
\frac{k_B T}{\hbar \omega_D} \!
\right\} ^{\! \frac{n}{2}} \!
\exp \!
\left( \!
- \frac{\hbar \omega_D}{k_B T}
\widetilde{\Delta }_T ^{
{\mbox{\scriptsize I}}({\mbox{\scriptsize II}})
} \!
\right) .~(n = 1,3, \cdots)
}
\label{Tdefinition}
\eeqa\\[-26pt]

In the opposite limit
$
T
\!\!\rightarrow\!\!
T_c ^{\mbox{\scriptsize I}}
$
($T_c$ for Case I)
the gap becomes very small,
$
[\det z_{12}]_T ^{1 / 2}
\!\!\rightarrow\!\! 1
$,
then
$
{\cal F}_{\! \Delta _T} ^{\uparrow }
\!\!\rightarrow\!\!
-
\Delta
N(0) V 
\mbox{arcsinh} \!
\left(
\hbar \omega_D / \Delta
\right) \!
/ 4
$
and
$
({\cal F}^{\uparrow }_{\! + \varepsilon_p T}
\!\!+\!\!
{\cal F}^{\uparrow }_{\! - \varepsilon_p T})
/ 2
\!\!\rightarrow\!\!
\varepsilon _p
/ 4
$
if we use the relations
(\ref{ResFepandFdp1})
and
(\ref{ResFepandFdp2}).
We have an approximate QP energy
$
\widetilde{\epsilon }_p ^{\uparrow \mbox{\scriptsize I}}
\!\simeq\!
[ 
\varepsilon _p ^2 
\!+\!
\left\{ \!
-\Delta \!
N \! (0) \! V \!
\mbox{arcsinh} \!
\left(
\hbar \omega_D \! / \! \Delta
\right) \!
\right\} ^2
]^{1/2} \!
/ \! 4
$
in which
the appearance of numerical factor $1 / 4$ should be become aware carefully.
This is because
two HB WFs
have different correlation structures
${\psi }_{2} \!=\! \pi$ and ${\psi }_{1} \!=\! 0$.
In such a case,
returning to the original form of the BCS gap equation but with
the modified QP energy
$\widetilde{\epsilon }_p ^{\uparrow \mbox{\scriptsize I}}$,
the thermal gap equation is expressed as 
$
1
\!\!=\!\!
V / 2
\sum_p \!
\left( 1 \!\!-\!\! 2 \widetilde{w}_{p} ^{\uparrow } \right) \!
/ \widetilde{\epsilon }_p ^{\uparrow \mbox{\scriptsize I}}
$
and leads to the integral form\\[-16pt]
\beqa
\begin{array}{c}
1
\!=\!
4 N(0) V \!\!
{\displaystyle
\int _0 ^{\hbar \omega_D} \!\!\! d \varepsilon
\frac{1}{\varepsilon }
\tanh \!
\left( \!
\frac{\varepsilon }{8k_B T ^{\mbox{\scriptsize I}}} \!
\right) .
}
\end{array}
\label{gapequation3}
\eeqa\\[-12pt]
We should emphasize that this form results from
the above numerical factor $1 / 4$.
Introduce a dimensionless variable
$
y_T ^{\mbox{\scriptsize I}}
\!\!\equiv\!\!
\varepsilon  / 8k_B T ^{\mbox{\scriptsize I}}
$,
its upper-value 
$
y ^{\mbox{\scriptsize I}}_{T_c ^{\mbox{\scriptsize I}}}
\!\!\equiv\!\!
\hbar \omega_D / 8k_B T_c ^{\mbox{\scriptsize I}}
$
and the Debye temperature
$\theta _D
\!\equiv\!
\hbar \omega_D / k_B
$.
Integrating R.H.S. of
(\ref{gapequation3})
 by parts, 
it is approximated as follows:\\[-20pt]
\beqa
\begin{array}{ll}
\!\!\!\!
{\displaystyle \frac{1}{4 N(0) V}}
&\!\!\simeq
\ln
y ^{\mbox{\scriptsize I}}_{T_c ^{\mbox{\scriptsize I}}}
\! - \!
{\displaystyle
\! \int _0 ^{\infty } \! \!\! dy
\ln y~\mbox{sech} ^2 y
} 
\! = \!
\ln
y ^{\mbox{\scriptsize I}}_{T_c ^{\mbox{\scriptsize I}}}
\! + \!
\ln \!
\left( \!
{\displaystyle \frac{4 e^C}{\pi } \!}
\right)
\! = \!
\ln \!
\left( \!
{\displaystyle
\frac{e^C}{2 \pi }
\frac{\hbar \omega_D}{k_B T_c ^{\mbox{\scriptsize I}}} \!
}
\right)
\!\equiv\!
\ln \!
\left( \!
{\displaystyle
\frac{\theta _D}{\widetilde{T}_c ^{\mbox{\scriptsize I}}} \!
}
\right) ,
\end{array}
\label{gapequation4}
\eeqa\\[-12pt]
where we have used the formula given in textbook
\cite{GraRyz.63}.
The number $C$ is
the Euler's constant 
$(C \!\simeq\! 0.5772)$ 
and
$e^C \!\simeq\! 1.781$.
Finally a small rearrangement yields\\[-18pt]
\beqa
\!\!
T_c ^{\mbox{\scriptsize I}}
\!=\!
 0.283 \theta _D
e ^{
- \frac{1}{4 N(0) V}
} .
\label{criticalTemp}
\eeqa\\[-20pt]
It should be compared with
the Eliashberg's formula
\cite{Eliashberg.61}
and
the usual HB's one
for $T_c$
\cite{Parks.69}\\[-18pt]
\beqa
T_c
\!=\!
 1.130 \theta _D
e ^{
- \frac{1}{N(0) V}
} .
\label{HBcriticalTemp}
\eeqa\\[-20pt]
The new formula
(\ref{criticalTemp})
gives a {\it high} critical-temperature,
e.g.,
$T_c ^{\mbox{\scriptsize I}} \!\!=\!\! 72.87 $K
for $N(0)V \!\!=\!\! 0.25$ and $\theta _D \!\!=\!\! 700 $K.
This $T_c ^{\mbox{\scriptsize I}}$ 
is in contrast to $T_c$
obtained by the famous HB formula
(\ref{HBcriticalTemp})
given by Rickayzen and Cohen in
\cite{Parks.69},
namely, $T_c \!\!=\!\! 14.49 $K for the same values
of $N(0)V$ and $\theta _D$.

From now we discuss
behaviour of the gap near $\!T_{\!c}$.$\!$
In the above the modified QP energy
$
\widetilde{\epsilon }
\!=\!\!
[\varepsilon ^2
\!+\!
\left\{ \!
-\Delta \!
N \!(0) V
\mbox{arcsinh} \!
\left(
\hbar \omega_D / \Delta
\right)
\right\}^2
]^{1/2}  \!
/ 4
$
plays a crucial role to boost the $T_{\!c}$
(\ref{criticalTemp})
comparing with $T_c \!\!=\!\! 14.49 $K by
(\ref{HBcriticalTemp}).
Note the numerical factor
$1 / 4$ in
$\widetilde{\epsilon }$.
First consider
$\Delta^{\mbox{\scriptsize I}} _T$
near $T_{\!c} ^{\mbox{\scriptsize I}}$.
Using this form of the modified QP energy,
the gap equation is roughly rewritten as\\[-18pt]
\beqa
\!\!\!\!\!\!\!\!
\begin{array}{c}
{\displaystyle
\frac{1}{4 N(0) V}
\!=\!
\frac{1}{4}
\int _0 ^{\hbar \omega_D} \! d \varepsilon
\frac{1}{\widetilde{\epsilon }}
\tanh \!
\left( \!
\frac{\widetilde{\epsilon }}{2k_B T} \!
\right) 
\!\simeq\! 
\int _0 ^{\hbar \omega_D} \! d \varepsilon
\frac{1}{\varepsilon }
\tanh \!
\left( \!
\frac{\varepsilon }{8k_B T} \!
\right)
}\\
\\[-14pt]
{\displaystyle
- \!
\left\{
\Delta _{\!T} \!
N(0) V \!
\! \cdot \! \mbox{arcsinh} \!
\left( \!
{\displaystyle \frac{\hbar \omega_D}{\Delta _{\!T}}} \!
\right) \!
\right\} ^{\!2}
\!\!
\int _0 ^{\hbar \omega_D} \!\! d \varepsilon
\left\{
\frac{1}{\varepsilon ^3}
\tanh \!
\left( \!
\frac{\varepsilon }{8k_B T} \!
\right)
\!-\!
\frac{1}{\varepsilon ^2}
\frac{1}{8k_B T}
\mbox{sech} ^{\!2} \!
\left( \!
\frac{\varepsilon }{8k_B T} \!
\right) \!
\right\} ,
}
\end{array}
\label{gapequation5}
\eeqa
\newpage
from which we obtain\\[-18pt]
\beqa
\begin{array}{ll}
& \!\!\!\!\!\!\!\!\!\! 
{\displaystyle
\frac{1}{4 N(0) V}
}
\!=\!
\ln \!
\left( \!
{\displaystyle
\frac{\hbar \omega_D}{k_B \widetilde{T}}
} \!
\right)
\!-\!
{\displaystyle \frac{7}{8 \pi ^2}}
\zeta (3) \!
\left( \!
{\displaystyle
\frac{2 \pi }{e^C}
} \!
\right) ^{\!2} \!\!
\left( \!
{\displaystyle
\frac{\hbar \omega_D}{k_B \widetilde{T}}
} \!
\right) ^{\!2} \!
\left\{ \!
N(0) V
x_T
\!\cdot\!
\mbox{arcsinh} \!
\left( \!
{\displaystyle
\frac{1}{x_T}
} \!
\right) \!
\right\}^2 \\
\\[-6pt]
& \!\!\!\!\!\!\!\!\!\!
\!\simeq\!
\ln \!
\left( \!
{\displaystyle
\frac{\hbar \omega_D}{k_B \widetilde{T}_c ^{\mbox{\scriptsize I}}}
} \!
\right)
\!+\!
{\displaystyle
\frac{\widetilde{T}_c ^{\mbox{\scriptsize I}} \!-\! \widetilde{T}}
{\widetilde{T}_c ^{\mbox{\scriptsize I}}}
}
\!-\!
{\displaystyle \frac{7}{8 \pi ^2}}
\zeta (3) \!
\left( \!
{\displaystyle
\frac{2 \pi }{e^C}
} \!
\right) ^{\!2} \!\!
\left( \!
1
\!-\!
{\displaystyle
\frac{\widetilde{T}_c ^{\mbox{\scriptsize I}} \!-\! \widetilde{T}}
{\widetilde{T}_c ^{\mbox{\scriptsize I}}}
} \!
\right) ^{\!\!-2} \!\!
\left( \!
{\displaystyle
\frac{\hbar \omega_D}{k_B \widetilde{T}_c ^{\mbox{\scriptsize I}}}
} \!
\right) ^{\!2} \!\!
\left\{ \!
N(0) V
x_T
\!\cdot\!
\mbox{arcsinh} \!
\left( \!
{\displaystyle
\frac{1}{x_T}
} \!
\right) \!
\right\}
^2 \! ,
\end{array}
\label{gapequation6}
\eeqa\\[-8pt]
where
$
\hbar \omega_D / k_B \widetilde{T}
\!\equiv\!
e^C / 2 \pi
\!\cdot\!
\hbar \omega_D / k_B T
$
and we have used the famous integral-formula
(\ref{integformula}).
Using
$
\mbox{arcsinh} \!
\left( \!
1 / x_T \!
\right)
\!\simeq\!
\ln
\left(
2 / x_T
\right)
\left(
0 \!<\! x_T\!<\! 1
\right)
$,
(\ref{gapequation4})
and
(\ref{gapequation6}),
we get $\Delta _T^{\mbox{\scriptsize I}}$
near $T_c ^{\mbox{\scriptsize I}}$ as\\[-18pt]
\beqa
{\displaystyle
\Delta _T^{\mbox{\scriptsize I}}
\!\simeq\!
2 \pi \sqrt{\frac{2}{7 \zeta (3)}}
\frac{k_B T_c ^{\mbox{\scriptsize I}}}{N(0) V} \!
\left(
1
-
\frac{T_c ^{\mbox{\scriptsize I}} - T}
{T_c ^{\mbox{\scriptsize I}}}
\right) \!
\sqrt{ \frac{T_c ^{\mbox{\scriptsize I}} - T}
{T_c ^{\mbox{\scriptsize I}}}}
} .
\label{gappnearTc}
\eeqa\\[-16pt]
Such a formula has been brought through the use of the modified QP energy
$\widetilde{\epsilon }$ which owes to a resonant feature of
the multi-band SC.
This new formula shows a more complicated
temperature-dependence of
$\Delta _T^{\mbox{\scriptsize I}}$
than the
$
\sqrt{T_c ^{\mbox{\scriptsize I}}}
\sqrt{T_c ^{\mbox{\scriptsize I}} \!-\! T}
$
dependence of
$\Delta _T^{\mbox{\scriptsize I}}$
presented
by equation (36.6) in textbook
\cite{AGD.65}
and by equations (16.32) and (16.33) in textbook
\cite{Abrikosov.88},
respectively.

Next,
for Case II,
$
({\cal F}^{\uparrow }_{\!\! + \varepsilon_p T}
\!+\!
{\cal F}^{\uparrow }_{\!\! - \varepsilon_p T})
$,
${\cal F}_{\!\! \Delta _T} ^{\uparrow }$ and
$\widetilde{\epsilon }_{\!p} ^{\uparrow }$
become infinite simultaneously in the limit
$\Delta _{\!T} \!\!\rightarrow\!\! 0$
$(x_{\!T} \!\!\rightarrow\!\! 0)$
due to the existence of
$1 \!\!-\!\! [\det z_{12}]_T ^{1 / 2}$
in the denominator.
Then mathematical handling for such a problem is too difficult and
therefore we can not easily get a formula
for $T_c ^{\mbox{\scriptsize II}}$
in an analytical way as we did in Case I.
Let us denote $T_c$ for Case II as
$T_c ^{\mbox{\scriptsize II}}$.
At $T \!\simeq\! T_c ^{\mbox{\scriptsize II}}$,
$\Delta _T^{\mbox{\scriptsize II}}$ almost vanishes
and 
$
1 \!-\! [\det z_{12}]_T ^{1 / 2}
\!\!\rightarrow\!\!
2 \pi N(0) \hbar \omega_{\!D} x_T
$.
Using the
${\cal F}_{\Delta _T} ^{\uparrow }$ expressed as
(\ref{ResFepandFdp2}),
we reach to the following asymptotic forms:
$
{\cal F}_{\! \Delta _T} ^{\uparrow }
\!\!\rightarrow\!\!
-
\left( 4 \pi N \! (0) \right)^{-1} \!
N \! (0) V 
\mbox{arcsinh}
\left(
\hbar \omega_{\!D} / \Delta _{\!T}
\right)
$
and
$
({\cal F}^{\uparrow }_{\! + \varepsilon_p T}
\!+\!
{\cal F}^{\uparrow }_{\! - \varepsilon_p T}) \!
/ 2
\!\!\rightarrow\!\!
\left( \! 4 \pi N \! (0) \! \right)^{-1} \!\!
\varepsilon _p / \! \Delta _T
$.
The QP energy
$
\widetilde{\epsilon }_p ^{\uparrow }
(
\!=\!
\{ \!
(
{\cal F}^{\uparrow }_{\! + \varepsilon_p T}
\!+\!
{\cal F}^{\uparrow }_{\! - \varepsilon_p T} \!
)^2 \!
/ 4
\!+\!
{\cal F}_{\! \Delta _T} ^{\uparrow 2} \!
\}^{\!1/2}
) \!
$
is approximately calculated as
$
\widetilde{\varepsilon }_p ^{\uparrow \mbox{\scriptsize II}}
\!\!\simeq\!\!
\left( 4 \pi N \! (0) \right)^{-1} \!
\varepsilon _p / \Delta _T
\left( 0 \! < \! \Delta _T \! \ll 1 \right)
$.
Here we discard the contribution from
${\cal F}_{\! \Delta _T} ^{\uparrow }$
comparing with the contribution from
$
({\cal F}^{\uparrow }_{\! + \varepsilon_p T}
\!+\!
{\cal F}^{\uparrow }_{\! - \varepsilon_p T})
/ 2$.
Returning again to the original form of the BCS gap equation but with
another modified QP energy
$\widetilde{\epsilon }_p ^{\uparrow \mbox{\scriptsize II}}$,
the thermal gap equation is obtained as
$
1
\!=\!
V / 2
\sum_p
\left( 1 \!-\! 2 \widetilde{w}_{p} ^{\uparrow } \right) \! /
\widetilde{\epsilon }_p ^{\uparrow \mbox{\scriptsize II}}
$
which also leads to the integral form\\[-14pt]
\beqa
\begin{array}{c}
1
\!=\!
{\displaystyle
N(0) V \!\! \int _0 ^{\hbar \omega_D} \!\! d \varepsilon
\frac{4 \pi N(0) \Delta _T ^{\mbox{\scriptsize II}}}{\varepsilon }
}
\tanh \!
\left(
{\displaystyle
\frac{\varepsilon }
{2k_B T ^{\mbox{\scriptsize II}} \!\cdot\! 4 \pi N(0)
\Delta _T ^{\mbox{\scriptsize II}}}
}
\right) .
\end{array}
\label{gapequation7}
\eeqa\\[-10pt]
We introduce the dimensionless variable
$
y_T ^{\mbox{\scriptsize II}}
\!\!\equiv\!\!
\varepsilon \! / \!\!
\left( \!
2 k_{\!B} \! T \!\!\cdot\! 4 \pi \! N \! (0) \!
\Delta _T ^{\mbox{\scriptsize II}} \!
\right)
$
and its upper-value
$
y ^{\mbox{\scriptsize II}}_{T ^{\mbox{\scriptsize II}}}
\!\!\equiv\!\!
\hbar \omega_{\!D} / \!
\left( \!
2 k_{\!B} \! T ^{\mbox{\scriptsize II}} \!\!\cdot\! 4 \pi \! N \! (0) \!
\Delta _T ^{\mbox{\scriptsize II}} \!
\right)
$.
Integrating
(\ref{gapequation7})
by parts
$(
\theta_D / k_B \! \widetilde{T}^{\mbox{\scriptsize II}}
\!\equiv\!
e^C \! / 2 \pi
\!\cdot\!
\hbar \omega_D \! / k_B \! T^{\mbox{\scriptsize II}}
)$,
we have\\[-18pt]
\beqa
\begin{array}{c}
{\displaystyle \frac{1}{4 N \! (0) V}}
{\displaystyle \frac{1}{\pi N \! (0) \Delta _T ^{\mbox{\scriptsize II}}}}
\!\simeq\!
\ln
y ^{\mbox{\scriptsize II}}_{T ^{\mbox{\scriptsize II}}}
\!+\!
\ln \!
\left( \!
{\displaystyle \frac{4 e^C}{\pi } \!}
\right)
\!=\!
\ln \!
\left( \!
{\displaystyle
\frac{1}{\pi N \! (0) \Delta _T ^{\mbox{\scriptsize II}}}
\frac{\theta _D}{\widetilde{T} ^{\mbox{\scriptsize II}}}
} \!
\right) ,
\end{array}
\label{gapequation8}
\eeqa\\[-16pt]
which reads\\[-22pt]
\beqa
\Delta _T ^{\mbox{\scriptsize II}}
=
{\displaystyle
\frac{\theta _D}{\pi \! N \! (0) \widetilde{T} ^{\mbox{\scriptsize II}}}
}
\exp
\left\{
{\displaystyle
-\frac{1}{4 N (0) V \pi N(0)
\Delta _T ^{\mbox{\scriptsize II}}}
}
\right\}
\approx
{\displaystyle
\frac{\theta _D}{\pi N (0) \widetilde{T} ^{\mbox{\scriptsize II}}}
}
\left(
1
-
{\displaystyle 
\frac{1}{4 N (0) V \pi N(0)
\Delta _T ^{\mbox{\scriptsize II}}}
}
\right) .
\label{eqdelT1}
\eeqa\\[-16pt]
From
(\ref{eqdelT1})
we obtain a quadratic equation for
$\Delta _T ^{\mbox{\scriptsize II}}$
very near $T_c$
and then we have a solution\\[-16pt]
\beqa
\Delta _T ^{\mbox{\scriptsize II}}
=
{\displaystyle
\frac{1}{\pi N(0)}
\frac{\theta _D}{\widetilde{T} ^{\mbox{\scriptsize II}}}
}
-
{\displaystyle \frac{1}{4 N (0) V \pi N(0)}} ,
\label{soldelT}
\eeqa\\[-16pt]
in which at
$
T ^{\mbox{\scriptsize II}}
=
T_c ^{\mbox{\scriptsize II}}
$,
the
$\Delta _T ^{\mbox{\scriptsize II}}$
vanishes.
Finally we can determine the critical temperature
$T_c ^{\mbox{\scriptsize II}}$
for Case II as\\[-28pt]
\beqa
T_c ^{\mbox{\scriptsize II}}
=
{\displaystyle \frac{2 e^C }{\pi }}
\theta _D
N (0) V
=
1.334
\theta _D
N (0) V .
\label{TcforII}
\eeqa\\[-18pt]
The simple formula
(\ref{TcforII})
gives a {\em high} critical temperature, e.g.,
$T_c ^{\mbox{\scriptsize II}}
\!=\!
$
198K for
$N (0) V
\!=\!
$
0.25
and
$\theta _D
\!=\!
$
700K.
Finally
$\Delta _T ^{\mbox{\scriptsize II}}$
near
$T_c ^{\mbox{\scriptsize II}}$
can be approximately obtained as\\[-14pt]
\beqa
\Delta _T ^{\mbox{\scriptsize II}}
\approx
-
{\displaystyle
\frac{e^C }{2 \pi }
\frac{1}{\pi N(0)}
\frac{\theta _D}{T_c ^{\mbox{\scriptsize II}}}
\frac{T - T_c ^{\mbox{\scriptsize II}}}
{T_c ^{\mbox{\scriptsize II}}}
} ,
\eeqa\\[-10pt]
which is linearly dependent on
$T \!\!-\!\! T_c ^{\mbox{\scriptsize II}}\!$.
$\!$It is very interesting that we could find
such a dependence of $\Delta _T^{\!\mbox{\scriptsize II}}$,
comparing with the usual dependence
$\!\sqrt{T \!\!-\!\! T_c ^{\mbox{\scriptsize II}}}\!$
of $\Delta _T^{\!\mbox{\scriptsize II}}$.
The numerical results
for
$N \! (0) \! V
\!\!=\!\!
$
0.25
and
$\theta _{\!D}
\!\!=\!\!
$
700K,
obtained from
(\ref{criticalTemp}), (\ref{HBcriticalTemp}) and (\ref{TcforII}),
are illustrated in Fig.\ref{figTc} below:
\vspace{0.1cm}
\begin{figure}[hbtp]
\begin{center}
\includegraphics[width=0.50\textwidth, height=0.35\textwidth]{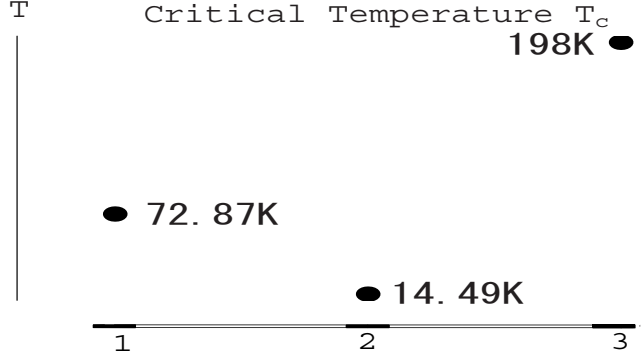}
\end{center}
\vspace{-1.5cm}
\caption{Critical Temperature $T_c$:
$~{\bf 1.}~T_c ^{\mbox{\scriptsize I}} \!=\! 72.87$K;
$~{\bf 2.}~T_c \!=\! 14.49$K;
$~{\bf 3.}~T_c ^{\mbox{\scriptsize II}} \!=\! 198$K
}
\label{figTc}
\end{figure}\\[-10pt]

\vspace{-0.3cm}

$\!\!\!\!$In the intermediate temperature region,
substituting
(\ref{DefinitionsofABandC2}) and (\ref{DefinitionsofABandC3})
into
(\ref{GapequationwithABandC}),
we have\\[-16pt]
\beqa
\!\!\!\!
\begin{array}{l}
\left[ \!
\left( \!
{\displaystyle
\frac{ e ^C }{\pi }
\frac{1}{1 \!\pm\! [\det z_{12}]_T ^{\frac{1}{2}}}
} \!\!
\right) ^{\!2} \!\!
x_{T} ^2 \!
\left\{ \!
N(0) V \mbox{arcsinh} \!
\left( \!
{\displaystyle \frac{1}{x_{T}}} \!\!
\right) \!
\right\}^{2} \!
\!-\!
{\displaystyle \frac{2 \pi ^2}{21 \zeta (3)}} \!
\ln \!
\left( \!
{\displaystyle
\frac{ e ^C }{\pi }
\frac{1}{1 \!\pm\!  [\det z_{12}]_T ^{\frac{1}{2}}}
\frac{\theta_D}{T}
}
\right) \!
\left( \!
{\displaystyle \frac{T}{\theta_D}} \!
\right) ^{\! 2}
\right] \\
\\[-14pt]
~\times
\left\{ \!
N(0) V \mbox{arcsinh}
\left( \!
{\displaystyle \frac{1}{x_{T}}} \!
\right)
\!-\!
\left( \!
1 \!-\! [\det z_{12}]_T ^{\frac{1}{2}}
\right) \!
\right\}
\!=\!
{\displaystyle \frac{2}{3}} \!
\left( \!
{\displaystyle
\frac{ e ^C }{\pi }
\frac{1}{1 \!\pm\!  [\det z_{12}]_T ^{\frac{1}{2}}}
} \!
\right) ^{\!2} \!
x_T ^2
[\det z_{12}]_T ^{\frac{1}{2}} . \\
\end{array}
\label{determinationofcriticaltemperature2}
\eeqa\\[-14pt]
A solution for equation
(\ref{determinationofcriticaltemperature2})
is classified into the following two cases:

Case I:The R.H.S. of
(\ref{determinationofcriticaltemperature2})
is approximated to be zero since
$0 \!<\! x_T \!\ll\! 1$,
from which we obtain an equation to determine
$\Delta_T$
for a given $T$ as\\[-18pt]
\beqa
\!\!
\begin{array}{c}
{\displaystyle
\frac{ e ^C }{\pi }
\frac{N(0) V }{1 \!+\! [\det z_{12}]_T ^{\frac{1}{2}}}
}
x_{T}
\mbox{arcsinh}
\left( \!
{\displaystyle \frac{2}{x_{T}}} \!
\right)
\!=\!
-
\sqrt{
{\displaystyle\frac{2 \pi ^2}{21 \zeta (3)}}
}
\sqrt{
\ln \!
\left( \!
{\displaystyle
\frac{ e ^C }{\pi }
\frac{1}{1 \!+\! [\det z_{12}]_T ^{\frac{1}{2}}}
\frac{\theta_D}{T}
} \!
\right)} \!
\left(
{\displaystyle
\frac{T}{\theta_D}
}
\right) .
\end{array}
\label{determinationofcriticaltemperature3}
\eeqa\\[-12pt]
Using the relation and the approximations\\[-18pt]
\beqa
\mbox{arcsinh}
\left( \!
{\displaystyle \frac{1}{x}} \!
\right)
\! = \!
\ln
\left( \!
{\displaystyle
\frac{1}{x} + \! \sqrt{1 \!+\! \frac{1}{x ^2}} \!
}
\right) ,~
{\displaystyle
\frac{1}{x} + \! \sqrt{1 \!+\! \frac{1}{x ^2}}
}
\!\approx\!
{\displaystyle
\frac{1}{x} + \! 1+ \!  \frac{1}{2}\frac{1}{x ^2}
} ,
\eeqa\\[-10pt]
and
$
e^x
\!\approx\!
1
\!+\!
x
\!+\!
\frac{1}{2}
x^2
$,
for
$[\det z_{12}]_T ^{1/2}
\!\approx\!
0.3
$
we have\\[-16pt]
\beqa
\!\!\!\!
\begin{array}{r}
x_{T}
\!+\!
{\displaystyle \frac{1}{2}}
\!=\!
-
{\displaystyle \frac{\sqrt{0.782}}{0.436 N(0) V}} \!
\sqrt{
\ln \!
\left( \! 0.436 {\displaystyle \frac{\theta_D}{T}} \!
\right)
} \!\!
\left( \!
{\displaystyle
\frac{T}{\theta_D} \!
}
\right) \!
x_{T}
\!+\!
{\displaystyle \frac{1}{2}}
{\displaystyle \frac{0.782}{\{0.436 N(0) V\}^2}} \!
\ln \!
\left( \!
0.436 {\displaystyle \frac{\theta_D}{T}} \!
\right) \!\!
\left( \!
{\displaystyle
\frac{T}{\theta_D}
} \!
\right)^{\!2} \! ,
\end{array}
\label{CaseI}
\eeqa\\[-10pt]
from which,
finally we have a very simple solution for
$x_T~(N(0) V \!=\! 0.25~\mbox{and}~\theta_{D} \!=\! 700\mbox{K})$
as\\[-16pt]
\beqa
\!\!\!\!\!\!
\begin{array}{l}
x_{T}
\!=\!
-
{\displaystyle \frac{1}{2}}
\left\{ \!
1
\!-\!
{\displaystyle
\frac{\sqrt{0.782}}{0.436 \!\!\times\!\! 0.25}
}
\sqrt{
\ln (0.436)
\!-\!
\ln \!
\left( \!\!
{\displaystyle
\frac{T}{700}
} \!\!
\right)
} \!\!
\left( \!\!
{\displaystyle
\frac{T}{700}
} \!\!
\right)  \!
\right\} .
\end{array}
\label{solCaseI}
\eeqa\\[-20pt]

Case II:Using
$
1 \!\!-\!\! [\det z_{12}]_T ^{1/2}
\!\!\approx\!\!
2 \pi \! N \! (0) \hbar \omega_{D} x_T
$,
from
(\ref{determinationofcriticaltemperature2})
we have\\[-18pt]
\beqa
\!\!\!\!\!\!\!\!\!\!\!\!\!\!\!\!
\begin{array}{c}
\left[ \!
(0.283)^2 \!
\left\{ \! N \! (0) V \! \right\}^{\!2} \!
\mbox{arcsinh}^2 \!\!
\left( \!\!
{\displaystyle
\frac{2}{x_{T}}
} \!\!
\right) \!
\!-\!
0.782 \!
\left\{ \! \pi \! N \! (0) \hbar \omega_D \! \right\}^{\!2} \!\!
\left[ \!
\ln \!\!
\left( \!\!
{\displaystyle
\frac{0.283}{2 \pi \! N \! (0) \hbar \omega_D}
} \!\!
\right) \!
\!-\!
\ln \!\!
\left( \!\!
{\displaystyle
\frac{T}{\theta_D} \!\!
}
\right) \!
\!+\!
\mbox{arcsinh} \!\!
\left( \!\!
{\displaystyle
\frac{2}{x_{T}}
} \!\!
\right) \!
\right] \!\!
\left( \!\!
{\displaystyle
\frac{T}{\theta_D}
} \!\!
\right)^{\!\!2} \!
\right] \\
\\[-14pt]
\!\times\!
\left\{ \!
N \! (0) V \!
\mbox{arcsinh} \!\!
\left( \!\!
{\displaystyle
\frac{2}{x_{T}}
} \!\!
\right)
\!-\!
\left( \!
1
\!-\!
[\det z_{12}]_T ^{\frac{1}{2}}
\right) \!
\right\}
\!=\!
-
{\displaystyle
\frac{2}{3}
}
(0.283)^2
[\det z_{12}]_T ^{\frac{1}{2}} .
\end{array}
\label{CaseII}
\eeqa\\[-10pt]
Expanding$\!$
(\ref{CaseII})
$\!$with respect to$\!$
$
\mbox{arcsinh} \!
\left( \!
2 \!/ \!x_{T} \!
\right)
$
$\!$and $\!$neglecting $\!$a$\!$ constant term$\!$
which is very small
for
$\![\det \! z_{12}]_T ^{1/2}
\!\!\approx\!
0.3
$
and for
$( \! N\!(0) V \!\!=\! 0.25,N\!(0) \hbar \omega_{\!D} \!=\! 0.01~\!\mbox{and}~\!
\theta_{\!D} \!=\! 700\mbox{K} \! )$,
(\ref{CaseII})
becomes to be a quadratic equation for
$
\mbox{arcsinh} \!
\left(
2 / x_T \!
\right)
$.
Finally we have the following solution for $x_T$:\\[-20pt]
\beqa
\!\!\!\!
\begin{array}{l}
x_T
\!=\!
2 /
\sinh \!\!
\left[
{ }^{^{^{^{^{^{^{^{^{^{^{.}}}}}}}}}}}
\!\!\!\!\!
{\displaystyle \frac{1}{2}} \!
\left\{
{\displaystyle
\frac{ \!
1
\!\!-\!\!
0.3
}
{ 0.25 }\!
\!+\!
\frac{ \!
0.782  \!
\left\{ \! \pi \!\!\times\!\!   0.01 \! \right\}^{\!2}}
{(0.283)^2 (0.25)^2}
} \!\!
\left( \!\!
{\displaystyle
\frac{T}{700}
} \!\!
\right)^{\!\!\!2} \!
\right\}
\right. \\
\\[-12pt]
\left.
+\!
\sqrt{ \!
{\displaystyle \frac{1}{4}} \!
\left\{ \!
{\displaystyle
\frac{ \!
1
\!\!-\!\!
0.3
}
{ 0.25 }
\!-\!
\frac{ \!
0.782  \!
\left\{ \! \pi\!\!\times\!\!   0.01 \! \right\}^{\!2}}
{(0.283)^2 (0.25)^2}
} \!\!
\left( \!\!
{\displaystyle
\frac{T}{700}
} \!\!
\right)^{\!\!\!2} \!
\right\}^{\!\!2}
\!\!\!+\!\!
{\displaystyle
\frac{
0.782 \!
\left\{ \! \pi \!\!\times\!\!   0.01 \! \right\}^{\!2}}
{(0.283)^2 (0.25)^3}
} \!\!
\left[ \!
0.25 \!
\left\{ \!
\ln \!\!
\left( \!\!
{\displaystyle
\frac{0.283}{2 \pi\!\!\times\!\!   0.01}
} \!\!
\right) \!
\!-\!
\ln \!\!
\left( \!\!
{\displaystyle
\frac{T}{700}
} \!\!
\right) \!\!
\right\} \!
\right] \!\!\!
\left( \!\!
{\displaystyle
\frac{T}{700}
} \!\!
\right)^{\!\!\!2} 
} \!\!\!
{ }^{^{^{^{^{^{^{^{^{^{^{.}}}}}}}}}}}
\right] \! .
\end{array}
\label{CaseII2}
\eeqa\\[-14pt]
We draw below the numerical results of the solutions for Case I and Case II.\\[-246pt]
\begin{figure}[hbtp]
\begin{minipage}{0.45\linewidth}
\vspace{7.9cm}
\includegraphics[width=1.0\textwidth,height=0.6\textwidth]{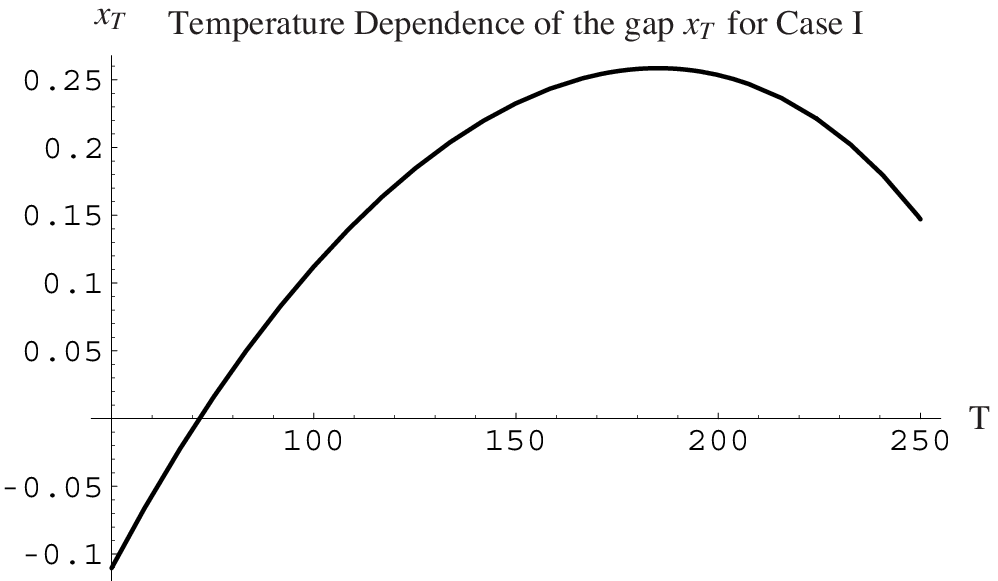}
\vspace{-0.90cm}
\caption{Temperature dependence of the gap,
Case I
for $[\det z_{12}]_T ^{1/2}
\!=\!
0.3
$, $N(0)V \!=\! 0.25$ and $\theta_D \!=\! 700$K.}
\label{fig:gap1}
\end{minipage}
\hfill
\begin{minipage}{0.45\linewidth}
\vspace{8.0cm}
\hspace{-0.19cm}
\includegraphics[width=1.0\textwidth,height=0.6\textwidth]{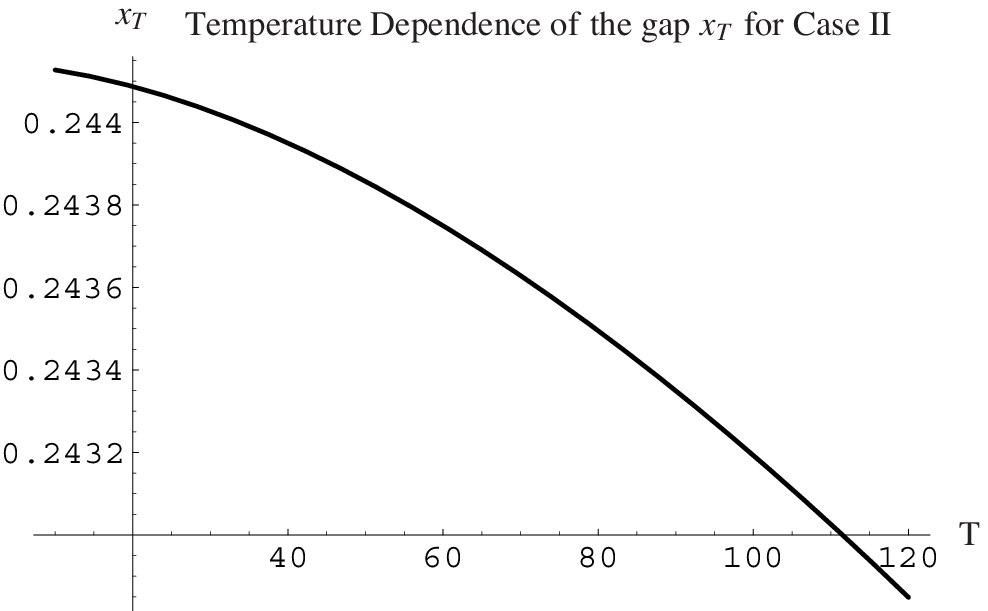}
\vspace{-0.93cm}
\caption{Temperature dependence of the gap,
Case II
for $[\det z_{12}]_T ^{1/2}
\!=\!
0.3
$, $N(0)V \!=\! 0.25,~N(0) \hbar \omega \!=\! 0.01$ and $\theta_D \!=\! 700$K.}
\label{fig:gap2}
\end{minipage}
\end{figure}

\vspace{-0.30cm}

The formula for Case I 
gives a {\it high} $T_c ^{\mbox{\scriptsize I}}$,
$\!$e.g.,$\!$
$T_c ^{\mbox{\scriptsize I}} \!\!=\!\! 72.87$K
for parameters $N(0) V \!\!=\!\! 0.25$ and 
$\theta _D \!\!=\!\! 700$K.
This is in contrast with $T_c$
of the usual HB formula 
giving $T_c \!\!=\!\! 14.49 $K for the same values,
$N(0)V \!\!=\!\! 0.25$ and $\theta _D \!\!=\!\! 700$.
The formula for Case II
gives also a {\it very high} $T_c ^{\mbox{\scriptsize II}}$,
e.g.,
$T_c ^{\mbox{\scriptsize II}} \!\!=\!\! 198 $K for the same values
of the parameters.
They are illustrated together in Fig.\ref{figTc}.
The temperature dependence of
gap near $T \!\!=\!\! 0$ and $T_c$
becomes more complicated than that of
the HB and 
Abrikosov's descriptions
\cite{AGD.65,Abrikosov.88}.
At intermediate temperature,
as shown in Figs. \ref{fig:gap1} and \ref{fig:gap2},
we have got the solutions of $\Delta_T$
for Cases I and II.
We assume
$[\det z_{12}]_T ^{1/2}
\!\!\approx\!
0.3
$ (Cases I and II)
and
$N \! (0) \hbar \omega_{\!D} \!\!=\!\! 0.01$ (Case II)
to acquire real solutions.
Anyway we could obtain really the solutions
$x_T \!\!=\!\! 0.030$ (Case I) and 0.243 (Case II)
for $T \!\!=\!\! 80$K.
The former has a negative gap below 72K.
To our great interest,
that value is almost equal to the $T_c$ given by
(\ref{criticalTemp}).
It, however, recovers a positive and small gap.
Further it increases as temperature rises up to around 190K
but shows vividly a decreasing tendency beyond around 200K.
In this sense the former is considerably good solution.
On the contrary,
the latter solution naturally decreases to 0.198 as temperature rises
up to around 2000K
but never vanishes.
This means the latter solution has no tendency approaching the $T_c$
given by
(\ref{TcforII}).
Much improvement of the above results should be possible
if the original equation
(\ref{determinationofcriticaltemperature2})
can be solved more accurately.

\newpage

%%%%%%%%%%%%%%%%%%%%%%%
%                                                               %
%  4   Summary and further perspectives   %
%                                                               %
%%%%%%%%%%%%%%%%%%%%%%%

\def\thesection{\arabic{section}}
\setcounter{equation}{0}
\renewcommand{\theequation}{\arabic{section}.\arabic{equation}}

\section{Summary and further perspectives}

~~~In this paper,
keeping an intimate connection with the usual BCS theory,
we have made an attempt at a Res-MF theoretical description
of the thermal behavior of the two-gap SC.
To show the predominance of the Res-HBT for
superconducting fermion systems with large quantum fluctuations
over the usual BCS and Eliashberg's theories, 
we have applied the Res-HBT to
the naive BCS Hamiltonian of singlet-pairing. 
We have obtained 
{\em gap equations}
within the framework of Res-HBA.
From the Res-FB operators
${\cal F}_1$ and ${\cal F}_2$
with {\em equal-gaps},
we have found the diagonalization condition for them,
which is essentially the same form as that of the BCS theory.
It leads to the self-consistent 
Res-HB gap equation, 
from which we could derive the present gap. 
Here we have concentrated 
on the derivation of the {\it Thermal Gap Equation}
with the use of the {\it thermal} Res-HBA.
From the {\it thermal} Res-FB operators
${\cal F}_{1T}$ and  ${\cal F}_{2T}$
with {\em equal-gaps}
we also have found the diagonalization condition, which is
just the same form of the condition at $T \!\!=\!\! 0$.
This reads the self-consistent 
Res-HB {\it Thermal Gap Equation}
and makes possible derivation of the new formulas to determine the $T_c$
and the gaps near $T \!\!=\!\! 0$ and $T_c$.

For {\em unequal two-gaps},
it is also possible to realize 
the above diagonalization condition
for Res-FB operators
${\cal F}_{rp}~(r \!=\! 1, 2)$.
Transforming by a unitary matrix
$\widehat{g}_{rp}$,
${\cal F}_{rp}$
is easily diagonalized.
Noticing the same correspondence as the correspondence in
(\ref{equivalenceforcosandsin}),
$\cos \theta _{rp} \!\!\Rightarrow\!\! \cos \widehat{\theta }_{rp}$
and
$\sin \theta _{rp} \!\!\Rightarrow\!\! \sin \widehat{\theta }_{rp}$,
we assume each diagonalization condition 
(\ref{equivalenceforsin3})
holds even in this case.
Then we obtain coupled equations through
a function of $\Delta_{1T}$ and $ \Delta_{2T}$
expressed as
\beqa
\begin{array}{c}
1
=
{\displaystyle
\frac{
{\displaystyle
\frac{\varepsilon _p ^2}
{(\varepsilon _p ^2 + \Delta ^2 _{r T}) ^{3/2}}
} 
\left( 
{\displaystyle
- \frac{2{\cal F}^{\uparrow }_{r \Delta _{T}}}{\Delta _{r T}}
}
\right)
\left(
1 - 2 \widetilde{w}^{\uparrow }_{r p}
\right)
}
{
{\displaystyle
\frac{\varepsilon _p}
{(\varepsilon _p ^2 + \Delta ^2 _{r T}) ^{3/2}}
} 
\left(
{\cal F}^{\uparrow }_{+ r \varepsilon p}
+
{\cal F}^{\uparrow }_{- r \varepsilon p} \!
\right) 
\left( 
1- 2 \widetilde{w}^{\uparrow }_{r p}
\right)
}
}, ~~
\widetilde{w}^{\uparrow }_{rp}
=
{\displaystyle
\frac{1}{1+
e^{{\displaystyle \beta \widetilde{\epsilon }_{rp}}}}
}, ~~
\Delta _{r T_c}
=
0 ,
\end{array}
\label{SCFconditionfortwogaps}
\eeqa
which reduces to
equation in R.H.S. of
(\ref{equivalenceforcosandsin})
if $\Delta_{1T} \!=\! \Delta_{2T}$.
The quantities
${\cal F}^{\uparrow }_{r, \Delta _{T}}$
and
${\cal F}^{\uparrow }_{r, \pm \varepsilon p}$
are given by the equations similar to
(5.9) in I
but with more complicated forms
of $\Delta_{1T}$ and $ \Delta_{2T}$.
For the time being,
as was done in the previous section
we here also use the function
$(\varepsilon _p ^2 \!+\! \Delta ^2 _{r T}) ^{3/2}$
by which we divide numerator and denominator, respectively,
in
(\ref{SCFconditionfortwogaps}).
After equating the numerator to the denominator and
using the relation
$
1 \!-\! 2 \widetilde{w}^{\uparrow }_{r p}
\!=\!
{\displaystyle
\tanh \!
\left(
\widetilde{\epsilon }_r / 2k_B T
\right)
}
$,
we sum up over $p$,
namely
integrate both sides of the equation over $\varepsilon$,
to achieve the optimized conditions.
Thus we obtain Res-HB coupled {\it Thermal Gap Equations}
and
reach our temporary goal of computing
thermal two-gaps.
Along such a strategy and method,
at the moment,
we will make a numerical analysis to demonstrate
the thermal behavior of two-gaps.

To describe a superconducting fermion system
and to approach such fundamental problems,
it is absolutely necessary to provide a rigorous thermal Res-HBA and MF approximation.
As mentioned in Introduction, we have the partition function as
$\mbox{Tr}(\!e^{-\beta H}\!)
\!\!=\!\!
2^{N \!-\! 1} \!\! \int \langle g |e^{-\beta H}| g \! \rangle dg$
and the projection operator $P$ onto the Res-HB subspace.
Then, the partition function in the Res-HB subspace is computed as
$\mbox{Tr}(\!Pe^{-\beta H}\!)$.
This can be calculated within the Res-HB subspace,
e.g,
by using the Laplace transform of
$e^{-\beta H}$
and the projection method.
The result leads to an infinite matrix continued fraction IMCF,
a concrete computation for which, however, is very difficult.
As a realistic problem,
it is better to seek for another possible and more practical way of
computing approximately the partition function and the Res-HB free energy
within the framework of the Res-MFT.
For this aim, it may be useful to introduce a quadratic Res-HB Hamiltonian
consisting of the Res-FB operators which satisfy
the Res-HB eigenvalue equations
$[{\cal F}_r u_r]_i = \epsilon _{ri} u_{ri}$.
This will be given elsewhere in a separate paper in a near future.

\newpage

%%%%%%%%%%
%                         %
%    Appendix     %
%                         %
%%%%%%%%%%

\vspace{1.0cm}
\leftline{\large{\bf Appendix}}
\appendix

%%%%%%%%%%%%%%%%%%%
%                                                   %
%  A  Proof of the equation (2.5)   %
%                                                   %
%%%%%%%%%%%%%%%%%%%

\def\thesection{\Alph{section}}
\setcounter{equation}{0}
\renewcommand{\theequation}{\Alph{section}.\arabic{equation}}
\section{Proof of the equation (\ref{traceformP})}

~~~~
The formula for the partition function in the Res-HB subspace,
(\ref{traceformP}),
is proved as follows:
Consider the whole Res-HB subspace\\[-16pt]
\beqa
\begin{array}{c}
|\Psi ^{\mbox{{\scriptsize Res}}(k)} \rangle
\!=\!
\sum _{t=1}^n
c_t ^{(k)}
|g _t \rangle , ~(k \!=\! 1, \cdots , n) 
\end{array}
\label{wholeRes-HBsubspace}
\eeqa
in which the Res-state with index
$k \!\!=\!\! 1$
and 
the Res-states with indices
$k \!\!=\!\! 2, \!\cdots\!, n$
stand for 
the Res-ground one and the Res-excited ones, respectively.
For each $k$ and $k^\prime$ state,
we regard the mixing coefficients $c_t ^{(k)*}$
and their products
$c_t ^{(k)*} \! c_{t^\prime } ^{(k^\prime)}$
as components of a column vector
$\left\{\!c ^{(k)*}\!\right\}$ and
matrix elements of a matrix
$\left\{\!
c ^{(k)*} \! c ^{(k^\prime) \mbox{{\scriptsize T}}}
\!\right\} $, respectively.
Then, we require the following ortho-normalization condition:\\[-16pt]
\beqa
\left.
\begin{array}{cc}
&\!\!\!
\langle
\Psi ^{\mbox{{\scriptsize Res}}(k)} |
\Psi ^{\mbox{{\scriptsize Res}}(k ^\prime)}
\rangle
\!=\!
\sum _{t,t^\prime=1}^n
c_t ^{(k)*} \! c_{t^\prime } ^{(k^\prime)}
S_{t t^\prime }
\!=\!
\mbox{Tr} \!
\left( \!
\left\{ \!
c ^{(k)*} \! c ^{(k^\prime) \mbox{{\scriptsize T}}} \!
\right\} \!
S ^{\mbox{{\scriptsize T}}} \!
\right)
\!=\!
0,~(k \!\neq\! k^\prime ) ,\\
\\[-6pt]
&\!\!\!
{\displaystyle \frac{1}{n}}
\sum _{k=1}^n
\langle
\Psi ^{\mbox{{\scriptsize Res}}(k)} |
\Psi  ^{\mbox{{\scriptsize Res}}(k)}
\rangle
\!=\!\!
{\displaystyle \frac{1}{n}}
\mbox{Tr} \!
\left(
\sum _{k=1}^n \!
\left\{ \!
c ^{(k)*} c ^{(k) \mbox{{\scriptsize T}}} \!
\right\} \!
S ^{\mbox{{\scriptsize T}}}
\right)
\!=\!
1 ,~
\langle
\Psi ^{\mbox{{\scriptsize Res}}(k)} | 
\Psi  ^{\mbox{{\scriptsize Res}}(k)}
\rangle
\!=\!
1,~
\forall k .
\end{array} \!
\right\}
\label{normalizationofeachRes-HBsubspace}
\eeqa
On the Res-WF 
$|\Psi ^{\mbox{{\scriptsize Res}}}\rangle$
we also demand the completeness condition\\[-16pt]
\beqa
\begin{array}{c}
{\displaystyle \frac{1}{n}}
\sum _{k = 1}^n
|
\Psi ^{\mbox{{\scriptsize Res}}(k)}
\rangle
\langle
\Psi ^{\mbox{{\scriptsize Res}}(k)}
|
=
{\displaystyle \frac{1}{n}}
\sum _{t,t^\prime=1}^n
|g _t \rangle
\sum _{k=1}^n
c_t ^{(k)} c_{t^\prime } ^{(k)*}
\langle g_{t^\prime } |
=
1 .
\end{array}
\label{completeness}
\eeqa
From 
(\ref{normalizationofeachRes-HBsubspace})
and
(\ref{completeness}),
we have an important relation\\[-16pt]
\beqa
\begin{array}{c}
\left\{
\sum _{k=1}^n 
c ^{(k)*} c ^{(k) \mbox{{\scriptsize T}}}
\right\}
_{t t^\prime }
\!=\!
(S ^{-1{\mbox{{\scriptsize T}}}})_{t {t^\prime }} .
\end{array}
\label{matrixofnbynproductccstar}
\eeqa
Using the definition of the projection operator $P$
(\ref{projectionoperator})
and
considering the meaning of the trace manupilation
in the present thermal Res-HBT,
the partition function in the Res-HB subspace
$\mbox{Tr} (Pe^{- \beta H}) $
is computed as\\[-16pt]
\beqa
\begin{array}{l}
\mbox{Tr} (Pe^{- \beta H})
\!=\!
\mbox{Tr} \!
\left( \!
\sum _{r,s=1}^n \!
|g _r \rangle (S^{-1})_{rs} \langle g _s | \!
e^{- \beta H}
\right) \\
\\[-8pt]
\!=\!
\sum _{k=1}^n \!
\sum _{t, t^\prime =1}^n
c_t ^{(k)*} \!
\langle g _t | \!
\sum _{r,s=1}^n \!
|g _r \rangle (S^{-1})_{rs} \langle g _s |
e^{- \beta H}
|g _{t^\prime } \rangle c_{t^\prime }^{(k)} \\ 
\\[-8pt]
\!=\!
\sum _{k=1}^n \!
\sum _{t, t^\prime =1}^n \!
c_t ^{(k)*} \! c_{t^\prime }^{(k)} \!
\sum _{r,s=1}^n \!
S_{tr}
(S^{-1})_{rs} \langle g _s |
e^{- \beta H}
|g _{t^\prime } \rangle %\\
%\\
\!=\!
\sum _{t, t^\prime =1}^n \!
\sum _{k=1}^n \!
c_t ^{(k)*} \! c_{t^\prime }^{(k)} \!
\langle g _t |
e^{- \beta H}
|g _{t^\prime } \rangle .
\end{array}
\label{traceform4}
\eeqa
Substituting (\ref{matrixofnbynproductccstar}) into 
(\ref{traceform4}),
thus, we obtain (\ref{traceformP}) exactly.
This is our desired result for the partition function.
This kind of trace formula is calculated within the Res-HB subspace
by using the Laplace transform of 
$e^{-\beta H}$
and the projection operator method
\cite{Naka.58,Zwan.60,Mori.65,Fulde.93}
which leads us to an infinite matrix continued fraction (IMCF).
In
(\ref{traceform4})
if we put the unit operator instead of
$e^{- \beta H}$
we get
$
\mbox{Tr} P
\!\!=\!\!
\sum _{t, t^\prime =1}^n \!
\sum _{k=1}^n \!
c_t ^{(k)*} \! c_{t^\prime }^{(k)} \!
\langle g _t |
e^{- \beta H}
|g _{t^\prime } \rangle
\!\!=\!\!
\sum _{t, t^\prime =1}^n 
(S^{-1})_{t^\prime t} (S)_{t t^\prime }
\!\!=\! n
$.
This means that
the entropy
$S_{\mbox{{\scriptsize Res}}}^{\mbox{\scriptsize thermalHB}}$
(See 
$F_{\mbox{{\scriptsize Res}}}^{\mbox{{\scriptsize thermalHB(2)}}}$
in
(\ref{ModificationofFreeEnergy}))
is at most $\ln n$. 
One expects that for sufficiently low temperatures the main
effect of temperature consists in inducing jumps from one resonating state
to another.
This effect may be described by the projection operator $P$.
In the case of $n \!=\! 2$,
$S_{\mbox{{\scriptsize Res}}}^{\mbox{\scriptsize thermalHB}}\!<\! \ln 2$.
This fact means, of course that the
extrapolations to higher temperatures may not be entirely reliable.
Nevertheless, we assume that by extrapolating the temperature behavior of
the gaps we may guess the critical temperatures.

As suggested in the last Section,
the partition function is also capable of computation
if we introduce a quadratic Res-HB Hamiltonian
consisting of the Res-FB operators which satisfy
the Res-HB eigenvalue equations
$[{\cal F}_r u_r]_i = \epsilon _{ri} u_{ri}$.
This may give another possible partition function
within the framework of the Res-MFT.

\newpage

%%%%%%%%%%%%%%%%%%%%%%
%                                                            %
%  B  Derivation of the equation (2.19)  %
%                                                            %
%%%%%%%%%%%%%%%%%%%%%%

\def\thesection{\Alph{section}}
\setcounter{equation}{0}
\renewcommand{\theequation}{\Alph{section}.\arabic{equation}}
\section{Derivation of the equation (\ref{solutionWrr})}

~~~~We here introduce the following Res-HB free energy
$F_{\mbox{{\scriptsize Res}}}
^{\mbox{{\scriptsize thermalHB}}}$
quite similar to
(\ref{densityform}).
We adopt
the thermal Lagrangian
$L_{\mbox{{\scriptsize Res}}}
^{\mbox{{\scriptsize thermalHB}}}$
(\ref{statisticalLagrangian})
without Lagrange multiplier term
$ E^{(k)}$
instead of the
$\langle H \rangle _{\mbox{{\scriptsize Res}}}$
but use
the entropy
$S_{\mbox{{\scriptsize Res}}}^{\mbox{\scriptsize thermalHB}}$,
namely,
multiplication of
$(-1/T)$
by
$F_{\mbox{{\scriptsize Res}}}^{\mbox{{\scriptsize thermalHB(2)}}}$
given right below,
which is expressed in terms of
the thermal HB density matrix
$W_{\mbox{{\scriptsize Res}}:rs}
^{\mbox{{\scriptsize thermal}}}$:\\[-14pt]
\beqa
\left.
\begin{array}{l}
F_{\mbox{{\scriptsize Res}}}
^{\mbox{{\scriptsize thermalHB}}}
=
F_{\mbox{{\scriptsize Res}}}
^{\mbox{{\scriptsize thermalHB(1)}}}
+
F_{\mbox{{\scriptsize Res}}}
^{\mbox{{\scriptsize thermalHB(2)}}} ,\\
\\[-8pt] 
F_{\mbox{{\scriptsize Res}}}
^{\mbox{{\scriptsize thermalHB(1)}}}
=
\sum_{k =1}^n \sum_{r,s =1} ^n
H[W_{\mbox{{\scriptsize Res}}:rs}^{\mbox{{\scriptsize thermal}}}]
\cdot [\det z_{rs}{^{\mbox{{\scriptsize thermal}}}]^\frac{1}{2}}
c_r ^{(k)*} c_s^{(k)} , \\
\\[-8pt]
F_{\mbox{{\scriptsize Res}}}
^{\mbox{{\scriptsize thermalHB(2)}}}
=
{\displaystyle \frac{1}{2}\frac{1}{\beta }}
\sum_{r, s =1} ^n
\mbox{Tr}
\left\{
W_{\mbox{{\scriptsize Res}}:rs}^{\mbox{{\scriptsize thermal}}}
\ln W_{\mbox{{\scriptsize Res}}:rs}^{\mbox{{\scriptsize thermal}}}
\right. \\
\\[-10pt]
\left.
~~~~~~~~~~~~~~~~~~~~~~~~~~~~~~~~~~~~~~~~~~~~~~~~~
\!+\!
(1_{2N} \!-\! W_{\mbox{{\scriptsize Res}}:rs}^{\mbox{{\scriptsize thermal}}})
\ln (1_{2N} \!-\! W_{\mbox{{\scriptsize Res}}:rs}^{\mbox{{\scriptsize thermal}}})
\right\} .
\end{array}
\right\}
\label{ModificationofFreeEnergy}
\eeqa
Multiplying the second equation of
(\ref{thermalRes-HBeigenvalueequation})
by
$W_{\mbox{{\scriptsize Res}}:rr}^{\mbox{{\scriptsize thermal}}}$
from the right
and using the explicit form of
${\cal K}_{\mbox{{\scriptsize Res}}:rr}
^{\mbox{{\scriptsize thermal}}(k)}$
obtained from
(\ref{thermalRes-HBequation})
and the idempotency relation
$
W_{\mbox{{\scriptsize Res}}:rr}^{\mbox{{\scriptsize thermal}}2}
\!=\!
W_{\mbox{{\scriptsize Res}}:rr}^{\mbox{{\scriptsize thermal}}}
$,
we can prove
the equivalence relation
and
the commutability relation\\[-12pt]
\beqa
\left.
\begin{array}{c}
\sum_{k =1}^n \sum_{s = 1} ^n
{\cal K}_{\mbox{{\scriptsize Res}}:rs}
^{\mbox{{\scriptsize thermal}}(k)}
c_r ^{(k)*} c_s^{(k)}
\equiv
{\cal F}_{\mbox{{\scriptsize Res}}:r}^{\mbox{{\scriptsize thermal}}}
W_{\mbox{{\scriptsize Res}}:rr}^{\mbox{{\scriptsize thermal}}}
\!-\!
W_{\mbox{{\scriptsize Res}}:rr}^{\mbox{{\scriptsize thermal}}}
{\cal F}_{\mbox{{\scriptsize Res}}:r}^{\mbox{{\scriptsize thermal}}}
W_{\mbox{{\scriptsize Res}}:rr}^{\mbox{{\scriptsize thermal}}},~ \\
\\[-6pt]
[{\cal F}_{\mbox{{\scriptsize Res}}:r}
^{\mbox{{\scriptsize thermal}}} ,
W_{\mbox{{\scriptsize Res}}:rr}^{\mbox{{\scriptsize thermal}}}] = 0 ,
\end{array}
\right\}
\label{modificationofWFW2}
\eeqa
which is identical to the thermal Res-HB equation
(\ref{thermalRes-HBequation}).
Further using the formulas
(\ref{statisticaldeltaWanddetz})
and
(\ref{modificationofWFW2}),
the direct variation of the Res-HB free energy is made
parallel to the variations carried out in
\cite{Fuku.88,NishiFuku.91}
as follows:\\[-14pt]
\beqa
\begin{array}{rl}
\delta F_{\mbox{{\scriptsize Res}}}
^{\mbox{{\scriptsize thermalHB(1)}}}
&\!\!\!\!
=
\sum_{r=1}^n
{\displaystyle \frac{1}{2}}\mbox{Tr}
\left\{
\sum_{k =1}^n \sum_{s=1}^n 
{\cal K}_{\mbox{{\scriptsize Res}}:rs}
^{\mbox{{\scriptsize thermal}}(k)}
c_r ^{(k)*} c_s^{(k)} u_r \delta u_r ^{\dagger }
\right\} \\
\\[-10pt]
&~~~~~~~~~~~~~~~~~~~~~+
\sum_{r=1}^n
{\displaystyle \frac{1}{2}}\mbox{Tr}
\left\{
\delta u_r u_r ^{\dagger }
\sum_{k =1}^n \sum_{s=1}^n
{\cal K}_{\mbox{{\scriptsize Res}}:rs}
^{\mbox{{\scriptsize thermal}}(k)\dagger }
c_r ^{(k)} c_s^{(k)*}
\right\} \\
\\[-12pt]
&\!\!\!\!
= 
\sum_{r=1}^n \!
{\displaystyle \frac{1}{2}}\mbox{Tr} \!
\left\{ \!
\left[
(
{\cal F}_{\mbox{{\scriptsize Res}}:r}^{\mbox{{\scriptsize thermal}}}
W_{\mbox{{\scriptsize Res}}:rr}^{\mbox{{\scriptsize thermal}}}
\!-\!
W_{\mbox{{\scriptsize Res}}:rr}^{\mbox{{\scriptsize thermal}}}
{\cal F}_{\mbox{{\scriptsize Res}}:r}^{\mbox{{\scriptsize thermal}}}
W_{\mbox{{\scriptsize Res}}:rr}^{\mbox{{\scriptsize thermal}}}
)
u_r \delta u_r ^{\dagger }
\right]
\right. \\
\\[-10pt]
&\left.
~~~~~~~~~~~~
\! + \!
\left[
\delta u_r u_r ^{\dagger }
(
W_{\mbox{{\scriptsize Res}}:rr}^{\mbox{{\scriptsize thermal}}}
{\cal F}_{\mbox{{\scriptsize Res}}:r}^{\mbox{{\scriptsize thermal}}}
\!-\!
W_{\mbox{{\scriptsize Res}}:rr}^{\mbox{{\scriptsize thermal}}}
{\cal F}_{\mbox{{\scriptsize Res}}:r}^{\mbox{{\scriptsize thermal}}}
W_{\mbox{{\scriptsize Res}}:rr}^{\mbox{{\scriptsize thermal}}}
)
\right] \!
\right\} \\
\\[-8pt]
&\!\!\!\!
=
\sum_{r=1}^n
{\displaystyle \frac{1}{2}}\mbox{Tr}
\left[
{\cal F}_{\mbox{{\scriptsize Res}}:r}
^{\mbox{{\scriptsize thermal}}}
(1_{2N} - W_{\mbox{{\scriptsize Res}}:rr}
^{\mbox{{\scriptsize thermal}}})
\delta W_{\mbox{{\scriptsize Res}}:rr}
^{\mbox{{\scriptsize thermal}}}
\right] ,
\end{array}
\label{variationoffreeenergy1}
\eeqa
\vspace{-0.2cm}
\beqa
\begin{array}{ll}
\!\!\!
\delta F_{\mbox{{\scriptsize Res}}}
^{\mbox{{\scriptsize thermalHB(2)}}}
&\!\!\!
\!=\!
{\displaystyle \frac{1}{2}\frac{1}{\beta }} \! \sum_{r=1}^n \! \mbox{Tr}
\left[ \!
\ln
\left\{ \!
W_{\mbox{{\scriptsize Res}}:rr}
^{\mbox{{\scriptsize thermal}}}
(1_{2N} \!-\! W_{\mbox{{\scriptsize Res}}:rr}
^{\mbox{{\scriptsize thermal}}})^{\!-1} \!
\right\} \!
(1_{2N} \!-\! W_{\mbox{{\scriptsize Res}}:rr}
^{\mbox{{\scriptsize thermal}}})
\delta W_{\mbox{{\scriptsize Res}}:rr}
^{\mbox{{\scriptsize thermal}}}
\right.\\
\\[-8pt]
&
\left.
\!\!\! + \!
\sum_{s=1}^{\prime~n} \!
(1_{2N} \!-\! W_{\mbox{{\scriptsize Res}}:rs}
^{\mbox{{\scriptsize thermal}}}) \!
\ln \!
\left\{ \!
W_{\mbox{{\scriptsize Res}}:rs}
^{\mbox{{\scriptsize thermal}}}
(1_{2N} \!-\! W_{\mbox{{\scriptsize Res}}:rs}
^{\mbox{{\scriptsize thermal}}})^{\!-1} \!
\right\} \!\!
W_{\mbox{{\scriptsize Res}}:rs}
^{\mbox{{\scriptsize thermal}}}
u_r \delta u_r ^{\dagger }
\!\!+\!
\mbox{h.c.} \!
\right] \! ,
\end{array}
\label{variationoffreeenergy2}
\eeqa
second line of
(\ref{variationoffreeenergy2})
has no contribution
since
$
(1_{2N} \!-\! W_{\mbox{{\scriptsize Res}}:rs}
^{\mbox{{\scriptsize thermal}}})
W_{\mbox{{\scriptsize Res}}:rs}
^{\mbox{{\scriptsize thermal}}}
\!=\! 0
$.
Then, the
variational\\[4pt] equation
$
\delta F_{\mbox{{\scriptsize Res}}}
^{\mbox{{\scriptsize thermalHB}}}
\!=\!
\delta F_{\mbox{{\scriptsize Res}}}
^{\mbox{{\scriptsize thermalHB(1)}}}
\!+\!
\delta F_{\mbox{{\scriptsize Res}}}
^{\mbox{{\scriptsize thermalHB(2)}}}
\!=\! 0
$
leads to
\beqa
\ln \!
\left\{ \!
W_{\mbox{{\scriptsize Res}}:rr}
^{\mbox{{\scriptsize thermal}}}
(1_{2N} \!-\! W_{\mbox{{\scriptsize Res}}:rr}
^{\mbox{{\scriptsize thermal}}})^{-1} \!
\right\}
\!=\! 
- \beta {\cal F}_{\mbox{{\scriptsize Res}}:r}
^{\mbox{{\scriptsize thermal}}} ,
\label{equilibriumcondition}
\eeqa
in which we have used the variational relations
$
\delta W_{\mbox{{\scriptsize Res}}:rr}
^{\mbox{{\scriptsize thermal}}}
\!=\!
u_r \delta u_r ^{\dagger } \!+\! \delta u_r u_r ^{\dagger }
$ 
and
$
\delta u_r ^{\dagger } u_r  \!+\! u_r ^{\dagger } \delta u_r  
\!=\! 0
$.\\[4pt]
From
(\ref{equilibriumcondition})
we get
$
~W_{\mbox{{\scriptsize Res}}:rr}
^{\mbox{{\scriptsize thermal}}}
(1_{2N} \!-\! W_{\mbox{{\scriptsize Res}}:rr}
^{\mbox{{\scriptsize thermal}}})^{-1}
\!=\!
\exp \{- \beta {\cal F}_{\mbox{{\scriptsize Res}}:r}
^{\mbox{{\scriptsize thermal}}} \} 
$.
Multiplication of the matrix\\[4pt]
$(1_{2N} - W_{\mbox{{\scriptsize Res}}:rr}
^{\mbox{{\scriptsize thermal}}})$ from the right
casts into
\beqa
\begin{array}{c}
W_{\mbox{{\scriptsize Res}}:rr}
^{\mbox{{\scriptsize thermal}}} 
=
\exp \{- \beta {\cal F}_{\mbox{{\scriptsize Res}}:r}
^{\mbox{{\scriptsize thermal}}} \}
(1_{2N} - W_{\mbox{{\scriptsize Res}}:rr}
^{\mbox{{\scriptsize thermal}}}) .
\end{array}
\label{relationWrr}
\eeqa
From
(\ref{relationWrr})
we can reach to the final goal of the desired equation
(\ref{solutionWrr}).

\newpage

%%%%%%%%%%%%%%%%%%%%%%%%%%%%%%%%%
%                                                                                             %
%  C  Calculations of sum p A p, sum  p B p and sum p C p   %
%                                                                                             %
%          at zero and intermediate temperature                          %
%                                                                                             %
%%%%%%%%%%%%%%%%%%%%%%%%%%%%%%%%%

\def\thesection{\Alph{section}}
\setcounter{equation}{0}
\renewcommand{\theequation}{\Alph{section}.\arabic{equation}}
\section{$\!\!$Calculations of $\sum _p \! A_p,\sum _p \! B_p$ and
$\sum _p \! C_p$ at zero and intermediate temperature}

~~First,
equation
(\ref{GapequationwithABandC})
 is shown to reduce to the Res-HB gap equation
(4.10) in I 
as $T \!\!\rightarrow\!\! 0$.
Using a variable $\varepsilon \!\!=\!\! \xi \Delta _T$ 
instead of $\varepsilon$,
summations
$\sum _p \! A_p,\sum _p \! B_p$ and $\sum _p \! C_p$
near $T \!\!=\!\! 0$
are computed as follows:\\[-16pt]
\beqa
\left.
\begin{array}{c}
{\displaystyle
\frac{\sum _p A_p}{2 N(0)}
=
\mbox{arcsinh}
\left(
\frac{1}{x_T}
\right)
- \frac{1}{\sqrt{1 + x ^2 _T}}
+
A(T) ,~~~
A(T)
=
- T ^{
{\mbox{\scriptsize I}}({\mbox{\scriptsize II}})
}_{\frac{3}{2}}
+
\cdots ,
} \\
\\[-14pt]
{\displaystyle
\frac{\Delta_T \sum _p B_p}{2 N(0)}
=
\arctan
\left(
\frac{1}{x_T}
\right)
+
B(T) ,~~~
B(T)
=
- T  ^{
{\mbox{\scriptsize I}}({\mbox{\scriptsize II}})
}_{\frac{1}{2}}
+ \frac{1}{2} T  ^{
{\mbox{\scriptsize I}}({\mbox{\scriptsize II}})
}_{\frac{3}{2}}
-
\cdots ,
} \\
\\[-14pt]
{\displaystyle
\frac{\Delta_T ^2 \sum _p C_p}{2 N(0)}
=
\frac{1}{\sqrt{1 + x ^2 _T}}
+
C(T) ,~~~
C(T)
=
- T  ^{
{\mbox{\scriptsize I}}({\mbox{\scriptsize II}})
}_{\frac{1}{2}}
+  T  ^{
{\mbox{\scriptsize I}}({\mbox{\scriptsize II}})
}_{\frac{3}{2}}
-
\cdots
},
\end{array}
\right\}
\label{TdependenceofABandC}
\eeqa\\[-12pt]
detailed calculation of which is given below.
With the use of the relations
(\ref{ResFepandFdp2})
and
$
\Delta _{\!T}/\varepsilon _p
\!=\!
$
$
-
2{\cal F}_{\! \Delta _T} ^{\uparrow }
/
({\cal F}^{\uparrow }_{\! + \varepsilon_p T}
\!+\!
{\cal F}^{\uparrow }_{\!\! - \varepsilon_p T}
)
$
which lead to
$\widetilde{\varepsilon }
\!=\! \sqrt{\xi ^2 \!+\! 1} \hbar \omega_D \widetilde{\Delta} _T^{\mbox{\scriptsize I(II)}}$,
$\!\sum _p \! A_p$ in
(\ref{DefinitionsofABandC})
is converted to\\[-16pt] 
\beqa
\begin{array}{c}
{\displaystyle
\frac{\sum _p A_p}{2 N(0)}
\!=\!
\int_{0}^{ \frac{1}{x}} \! d\xi
\frac
{\xi ^2}
{(\xi ^2 \!+\! 1) ^{\frac{3}{2}}}
\!-\!
2 \!
\int_{0}^{ \frac{1}{x}} \! d\xi
\frac
{\xi ^2}
{(\xi ^2 \!+\! 1) ^{\frac{3}{2}}}
\frac{1}{1 \!+\! e ^{{\displaystyle {\bf d}}^{\mbox{\scriptsize I(II)}} 
{\displaystyle \sqrt{\xi ^2 \!+\! 1}}}}
} ,~
{\displaystyle
{\bf d} ^{\mbox{\scriptsize I(II)}} 
\!\equiv\!
\frac{\hbar \omega_D}{k_B T}
\widetilde{\Delta}_T ^{\mbox{\scriptsize I(II)}}
} ,
\end{array}
\label{CalsTempgap}
\eeqa\\[-10pt]
where
${\bf d} ^{\mbox{\scriptsize I(II)}} \gg 1$
for large $\hbar \omega_D$
and
for Case I and Case II,
$\widetilde{\Delta }_T  ^{\mbox{\scriptsize I(II)}}$
is defined as
\\[-10pt] 
\beq
\widetilde{\Delta }_T  ^{\mbox{\scriptsize I(II)}}
\! \equiv \!
{\displaystyle
\frac{1}{2}}
x_T
\left\{
N(0) V
\! \cdot \! \mbox{arcsinh}
\left( \!
\frac{1}{x_T} \!
\right)
\! +(-)
[\det z_{12}]^{\frac{1}{2}} \!
\right\}
\! \cdot \!
{\displaystyle \frac{1}
{1 \! +(-) [\det z_{12}]^{\frac{1}{2}}}} .
\label{Deltatilde}
\eeq\\[-10pt]
Introducing a new variable $y =  \sqrt{\xi ^2 + 1}$,
(\ref{CalsTempgap}) 
is integrated partly and approximated to be\\[-18pt]
\beqa
\!\!\!\!
\begin{array}{c}
{\displaystyle
\frac{\sum _p A_p}{2 N(0)}
\!\simeq\!
\mbox{arcsinh}
\left( \!
\frac{1}{x_T} \!
\right)
\!-\!
\frac{1}{\sqrt{1 \!+\! x_T ^2}}
\!-\!
2 \!\!
\int_{1}^{\infty } \!\! dy
\frac{1}{\sqrt{y ^2 \!-\! 1}}
e^{-{\displaystyle {\bf d}}^{\mbox{\scriptsize I(II)}}
{\displaystyle y}}
}
\!+\!
2 \!
{\displaystyle 
\int_{1}^{\infty } \!\! dy
\frac{1}{y ^2}
\frac{1}{\sqrt{y ^2 \!-\! 1}}
e^{-{\displaystyle {\bf d}}^{\mbox{\scriptsize I(II)}}
{\displaystyle y}}.
}
\end{array}
\label{CalsTempgapA}
\eeqa\\[-24pt]

Similarly, we get approximate formulas for
$\sum _p B_p$ and $\sum _p C_p$
as\\[-10pt]
\beq
{\displaystyle
\frac{\Delta_T \sum _p B_p}{2 N(0)}
\simeq
\arctan
\left(
\frac{1}{x_T}
\right)
- 2
\int_{1}^{\infty } dy
\frac{1}{y }\frac{1}{\sqrt{y ^2 - 1}}~
e^{-{\displaystyle {\bf d}}^{\mbox{\scriptsize I(II)}}
{\displaystyle y}},
}
\label{CalsTempgapB}
\eeq
\vspace{-0.2cm}
\beq
{\displaystyle
\frac{\Delta_T ^2 \sum _p C_p}{2 N(0)}
\simeq
\frac{1}{\sqrt{1 + x_T ^2}}
- 2
\int_{1}^{\infty } dy
\frac{1}{y ^2}\frac{1}{\sqrt{y ^2 - 1}}~
e^{-{\displaystyle {\bf d}}^{\mbox{\scriptsize I(II)}}
{\displaystyle y}}.
}
\label{CalsTempgapC}
\eeq\\[-10pt]
To carry out integral calculations in 
(\ref{CalsTempgapA}) 
$\sim$
(\ref{CalsTempgapC}),
it is convenient to use 
an integral representation of Bessel function
\cite{Watson.53}. 
The Bessel function of order $\nu$ is represented as\\[-10pt]
\beq
K_\nu (z)
=
\frac{\sqrt{\pi }
\left(
\frac{z}{2}
\right) ^\nu }
{\Gamma
\left(
\nu +  \frac{1}{2}
\right)}
\int_{1}^{\infty } dy
(y ^2 - 1) ^{\nu -  \frac{1}{2}}e^{-zy} .
\label{BesselfunctionKn}
\eeq\\[-10pt]
Then, the integral form of the Bessel function of order $0$ 
and
its exact result are given by\\[-10pt]
\beq
\begin{array}{c}
{\displaystyle
K_0 (z)
=
\frac{\sqrt{\pi }}
{\Gamma
\left(
\frac{1}{2}
\right)}
\int_{1}^{\infty } dy
\frac{1}{\sqrt{y ^2 - 1}}~e^{-zy}
= \sqrt{\frac{\pi }{2z}}e^{-z},
}~
(\Gamma
\left(
\frac{1}{2}
\right)
=
\sqrt{\pi }).
\end{array}
\label{BesselfunctionK0}
\eeq\\[-10pt]
Using (\ref{BesselfunctionK0}),
integral calculations of
(\ref{CalsTempgapA})
$\sim$
(\ref{CalsTempgapC})
are made in the following ways:\\[-16pt]
\beqa
\begin{array}{c}
{\displaystyle
\int_{1}^{\infty } dy
\frac{1}{y}\frac{1}{\sqrt{y ^2 - 1}}~e^{-{\displaystyle{\bf d}y}}
=
\int_{1}^{\infty } dy
\frac{1}{\sqrt{y ^2 - 1}}
\int_{{\displaystyle{\bf d}}}^{\infty } dz
e^{-zy}
=
\int_{{\displaystyle{\bf d}}}^{\infty } dz K_0 (z)
} \\
\\[-12pt]
{\displaystyle
~~~~~~~~~~~~~~~~~~~~~=
\sqrt{\frac{\pi }{2}}
\left(
{\bf d} ^{-\frac{1}{2}} - \frac{1}{2}{\bf d} ^{-\frac{3}{2}}
+ \frac{3}{4}{\bf d} ^{-\frac{5}{2}}
- \cdots
\right)
e^{-{\displaystyle{\bf d}}}
},
\end{array}
\label{integralformula1forBesselfunc}
\eeqa
\vspace{-0.5cm}
\beqa
\begin{array}{c}
{\displaystyle
\int_{1}^{\infty } dy
\frac{1}{y ^2}\frac{1}{\sqrt{y ^2 - 1}}~e^{-{\displaystyle{\bf d}y}}
=
\int_{1}^{\infty } dy
\frac{1}{y }
\frac{1}{\sqrt{y ^2 - 1}}
\int_{{\displaystyle{\bf d}}}^{\infty } dw
e^{-wy}
=
\int_{{\displaystyle{\bf d}}}^{\infty } dw
\int_{w}^{\infty } dz K_0 (z)
} \\
\\[-12pt]
~~~~~~~~=
{\displaystyle
\sqrt{\frac{\pi }{2}}
\left(
{\bf d} ^{-\frac{1}{2}} - {\bf d} ^{-\frac{3}{2}}
+ \frac{9}{4}{\bf d} ^{-\frac{5}{2}}
- \cdots
\right)
e^{-{\displaystyle{\bf d}}}
}.
\end{array}
\label{integralformula2forBesselfunc}
\eeqa\\[-10pt]
As a result, we obtain the approximation
for $A(T)$, $B(T)$ and $C(T)$ near $T=0$ as\\[-18pt]
\beqa
A(T) \simeq 0,~~~
B(T)
=
C(T)
\simeq
- T  ^{
{\mbox{\scriptsize I}}({\mbox{\scriptsize II}})
}_{\frac{1}{2}}.
\label{Approxmations2}
\eeqa\\[-18pt]
Further in the above near $T \!=\! 0$
we make the following approximations:\\[-16pt]
\beqa
\!\!\!\!
\left.
\begin{array}{c}
{\displaystyle
\arctan \!
\left( \!
\frac{1}{x_T} \!
\right)
\simeq
\frac{\pi }{2} \!-\! x_T,~~
\frac{1}{\sqrt{1 \!+\! x ^2 _T}}
\!\simeq\!
1 - x_0 x_T,~~
(0 \!<\! x_0 \!\ll\! 1)
}\\
\\[-12pt]
{\displaystyle
\mbox{arcsinh} \!
\left( \!
\frac{1}{x_T} \!
\right)
\simeq
\mbox{arcsinh} \!
\left( \!
\frac{1}{x_0} \!
\right)
-
\frac{1}{x_0}
(x_T - x_0),~~
[\det z_{12}]_T ^{\frac{1}{2}}
\simeq
[\det z_{12}]^{\frac{1}{2}}_{T=0} 
} .
\end{array} \!
\right\}
\label{Approxmations}
\eeqa\\[-22pt]

Next, let us introduce a new variable $y$ by
$
\varepsilon
\!\!=\!\!
4
( \!
1 \!\pm\! [\det \! z_{12}]_T ^{1/2}
)
k_B T \! y
$
and
quantities
$
\widetilde{x}_T
\!\!=\!\!
\widetilde{\Delta }_T \! / \! \hbar \omega_{\!D}
$
and
$
y_T  ^{(\pm)}
\!\!=\!\!
\sqrt{\varepsilon ^2 \!+\! \widetilde{\Delta }_T ^2}
\{
4 ( 1 \!\pm\! [\det z_{12}]_T ^{1/2} )
\}^{\!-1}
\hbar \omega_{\!D} / \! k_B T
$
where
$
\widetilde{\Delta }_T
\!\equiv\!
\Delta _T N(0) V
\mbox{arcsinh}
\left(
\hbar \omega_{\!D} / \! \Delta_T
\right)
$.
In intermediate temperature region
the modified QP energy $\widetilde{\varepsilon }$
is approximated as
$
\widetilde{\varepsilon } ^{(\pm)}
\!=\!
\sqrt{\varepsilon ^2 \!+\! \widetilde{\Delta }_T ^2}
\{
2 ( 1 \!\pm\! [\det z_{12}]_T ^{1/2} )
\}^{\!-1}
$.
If
$\varepsilon \!\!\gg\!\! \Delta_T$,
$\sum _p \! A_p, \sum _p \! B_p$ and $\sum _p \! C_p$ in
(\ref{DefinitionsofABandC})
are recast to the following integrals up to
$\widetilde{\Delta }_T$:\\[-14pt]
\beqa
\begin{array}{c}
{\displaystyle
\frac{\sum _p \! A_p}{2 N(0)}
}
\simeq
{\displaystyle
\int _{0} ^{\hbar \omega_D}
\!\!\!\! d \varepsilon 
\frac
{1}
{\sqrt{\varepsilon  ^2 \!+\! \widetilde{\Delta }_T ^2 }}
\tanh \!
\left( \!\!
\frac{\widetilde{\varepsilon } ^{(\pm)}}{2k_B T} \!\!
\right)
-
\int _{0} ^{\hbar \omega_D}
\!\!\!\! d \varepsilon
\frac
{ \widetilde{\Delta }_T ^2}
{\varepsilon ^2 \sqrt{\varepsilon ^2 \!+\! \widetilde{\Delta }_T ^2 }}
\tanh \!
\left( \!\!
\frac{\widetilde{\varepsilon } ^{(\pm)}}{2k_B T} \!\!
\right)
} \\
\\[-6pt]
\!\!\!\!\!\!
=
{\displaystyle
\int _{0} ^{y_T ^{(\pm)}}
\!\!\! dy
}
\left\{ \!
{\displaystyle
1
-
\frac{
\left( \!
y_T ^{(\pm)}
\widetilde{x}_T \!
\right) ^2}{y ^2}
} \!
\right\} \!
{\displaystyle
\frac
{1}
{
\sqrt{y  ^2 \!\!+\!\!
\left( \!
y_T ^{(\pm)}
\widetilde{x}_T \!
\right) ^2 }
}
} \!
\tanh \!
\left[ \!
\sqrt{y ^2 \!\!+\!\! 
\left( \!
y_T ^{(\pm)}
\widetilde{x}_T \!
\right) ^2 }
\right] ,
\end{array}
\label{CalofA}
\eeqa
\vspace{-0.2cm}
\beqa
\begin{array}{c}
\!\!\!\!
{\displaystyle
\frac{\widetilde{\Delta }_{\!T} \!\! \sum _p \! B_p}{N(0)}
}
\simeq
{\displaystyle
\int _{-\hbar \omega_D} ^{\hbar \omega_D} \!\!
\!\! d \varepsilon \!
}
{\displaystyle
\frac{\widetilde{\Delta }_T}
{\varepsilon \! \sqrt{\varepsilon ^2
\!\!+\!\!
\widetilde{\Delta }_T ^2}}
} \!
\tanh \!
\left( \!\!
{\displaystyle
\frac{\widetilde{\varepsilon } ^{(\pm)}}{2k_B T}
} \!\!
\right)
{\displaystyle
-
{\displaystyle \frac{1}{2}} \!
\int _{-\hbar \omega_D} ^{\hbar \omega_D} \!\!
\!\! d \varepsilon \!
}
{\displaystyle
\frac{\widetilde{\Delta }_T ^3}
{\varepsilon ^3 \! \sqrt{\varepsilon ^2
\!\!+\!\!
\widetilde{\Delta }_{\!T} ^2}}
} \!
\tanh \!
\left( \!\!
{\displaystyle
\frac{\widetilde{\varepsilon } ^{(\pm)}}{2k_B T}
} \!\!
\right)
= 0,
\end{array}
\label{CalofB}
\eeqa
\vspace{-0.2cm}
\beqa
\!\!\!\!\!\!
\begin{array}{rl}
&
{\displaystyle
\frac{ \widetilde{\Delta }_T ^2 \! \sum _p \! C_p}{2 N(0)}
}
\simeq
{\displaystyle
\int _{0} ^{\hbar \omega_D}
\!\!\!\!\! d \varepsilon
}
{\displaystyle
\frac{\widetilde{\Delta }_T ^2}
{\varepsilon ^2 \! \sqrt{\varepsilon  ^2
\!\! + \!\!
\widetilde{\Delta }_T ^2}} \!
\tanh \!
\left( \!\!
\frac{\widetilde{\varepsilon }^{(\pm)}}{2k_B T} \!\!
\right)
-
\int _{0} ^{\hbar \omega_D}
\!\!\!\!\! d \varepsilon
\frac
{ \widetilde{\Delta }_T ^4}
{\varepsilon  ^4 \! \sqrt{\varepsilon ^2 \!\! + \!\! \widetilde{\Delta }_T ^2}} \!
\tanh \!
\left( \!\!
\frac{\widetilde{\varepsilon }^{(\pm)}}{2k_B T} \!\!
\right)
}\\
\\[-6pt]
&=
{\displaystyle
\left( \!
y_T ^{(\pm)}
\widetilde{x}_T \!
\right) ^{\!2} \!
\int _{0} ^{y_T ^{\mbox{\scriptsize II}}}
\!\!\! dy \!
}
\left\{ \!
{\displaystyle
1
-
\frac{\left(
y_T ^{(\pm)}
\widetilde{x}_T \!
\right) ^{\!2}}{y ^2}
} \!
\right\}
{\displaystyle
\frac{1}
{y ^2 \! \sqrt{y  ^2 \!\! + \!\!
\left(
y_T ^{(\pm)}
\widetilde{x}_T \!
\right) ^{\!2} }
}
}
\tanh \!
\left[ \!
\sqrt{y  ^2 \!\! + \!\!
\left(
y_T ^{(\pm)}
\widetilde{x}_T \!
\right) ^{\!2} }
\right] .
\end{array}
\label{CalofC}
\eeqa%\\[-10pt]
Further equations
(\ref{CalofA}) and (\ref{CalofC})
are approximately computed, respectively, as\\[-14pt]
\beqa
\!\!\!\!\!\!\!\!
\left.
\begin{array}{c}
{\displaystyle
\frac{\sum _p \! A_p}{2 N(0)}
=
\int _{0} ^{\hbar \omega_D} \!\!\!\!\! d \varepsilon
\frac
{\varepsilon ^2}
{
\left( \!
\varepsilon  ^2
\!\!+\!\!
\widetilde{\Delta }_T ^2 \!
\right)^{\frac{3}{2}}
}
\tanh \!
\left( \!\!
\frac{\widetilde{\varepsilon } ^{(\pm)}}{2k_B T} \!\!
\right)
}
{\displaystyle
\simeq
\ln \!
\left( \!
\frac{4 e^C}{\pi }
y_T ^{(\pm)} \!
\right)
-
\frac{21}{2 \pi ^2}
\zeta (3) \!
\left( \!
y_T ^{(\pm)}
\widetilde{x}_T \!
\right) ^{\!2}
}, \\
\\[-10pt]
{\displaystyle
\frac{\widetilde{\Delta }_T ^2 \! \sum _p \! C_p}{2 N(0)}
=
\int _{0} ^{\hbar \omega_D} \!\!\!\!\! d \varepsilon
\frac
{1}
{
\left( \!
\varepsilon ^2
\!\!+\!\!
\widetilde{\Delta }_T ^2 \!
\right)^{\!\frac{3}{2}}
} \!
\tanh \!
\left( \!\!
\frac{\widetilde{\varepsilon }^{(\pm)}}{2k_B T} \!\!
\right)
}
{\displaystyle
\simeq
\frac{7}{\pi ^2}
\zeta (3) \!
\left( \!
y_T ^{\mbox{\scriptsize II}}
x_T \!
\right) ^{\!2}
} \! .
\end{array}
\right\}
\label{DefinitionsofABandC2}
\eeqa\\[-12pt]
Taking only a leading term,
finally terms $A_p$ and $C_p$ in
(\ref{DefinitionsofABandC2})
are approximated to be\\[-12pt]
\beqa
\left.
\begin{array}{l}
{\displaystyle
\frac{\sum _p \! A_p}{2 N(0)}
\simeq
\ln \!
\left( \!
\frac{ e ^C }{\pi }
\frac{1}{1 \pm [\det z_{12}]_T ^{\frac{1}{2}}}
\frac{\theta_D}{T} \!
\right)
-
\alpha _T ^{(\pm)}
} ,~
{\displaystyle
\frac{\widetilde{\Delta }_T ^2 \! \sum _p \! C_p}{2 N(0)}
\simeq
\frac{2}{3}
\alpha_T ^{(\pm)} \!
} , \\
\\[-4pt]
{\displaystyle
\alpha_T ^{(\pm)}
\equiv
\frac{21 \zeta (3)}{2 \pi ^2} \!
\left( \!
\frac{ e ^C }{\pi }
\frac{1}{1 \pm [\det z_{12}]_T ^{\frac{1}{2}}} \!
\right) ^{\!2} \!
\left( \!
\frac{\widetilde{\Delta }_T}{k_B T} \!
\right) ^{\!2} \!
} .
\end{array} \!\!
\right\}
\label{DefinitionsofABandC3}
\eeqa\\[-6pt]
To derive
(\ref{DefinitionsofABandC2})
we give a integral formula\\[-4pt]
\beq
\int _{0} ^\infty
\!\!\! dy
\left\{
\frac
{1}{y ^3}
\tanh y
-\!
\frac{1}{y ^2}
\mbox{sech} ^2 y
\right\} 
\!=\!
\frac{7}{\pi ^2}\zeta (3) ,
\label{integformula}
\eeq\\[1pt]
which can be derived
by using the famous mathematical formulas
\cite{GraRyz.63}\\[-14pt]
\beqa
\!\!\!\!
\begin{array}{c}
{\displaystyle
\frac{1}{y}
\tanh y
}
\!=\!
8 \! \sum_{m \!=\! 1}^{\infty } \!
{\displaystyle
\frac{1}{(2m \!\!-\!\! 1)^2 \pi ^2 \!\!+\!\! 4 y ^2}
} , ~
{\displaystyle
\frac{1}{y ^3}
\tanh y
-\!
\frac{1}{y ^2}
\mbox{sech} ^2 y
}
\!=\!
64 \! \sum_{m \!=\! 1}^{\infty } \!
{\displaystyle
\frac{1}
{
\left\{
(2m \!\!-\!\! 1)^2 \pi ^2 \!\!+\!\! 4 y ^2
\right\} ^2
}
}.
\end{array}
\label{mathformulas}
\eeqa
Adopting a new integral variable
$
y
\!=\!
(2m \!\!-\!\! 1) \pi /2
\!\cdot\!
\tan \theta
$,
an integral of the second formula in
(\ref{mathformulas}) is easily carried out for
$\hbar \omega_D \gg 1$
as\\[-16pt]
\beqa
\begin{array}{c}
{\displaystyle
64 \!
\int _{0} ^{y_T ^{\mbox{\scriptsize II}}\rightarrow \infty }
}
\!\!\! d{y} \!
\sum _{m \!=\! 1}^{\infty } \!
{\displaystyle
\frac{1}
{
\left\{
(2m \!\!-\!\! 1)^2 \pi ^2 \!\!+\!\! 4 y ^2
\right\} ^2
}
}
\!=\!
32
\sum _{m \!=\! 1}^{\infty } \!
{\displaystyle
\frac{1}
{(2m \!\!-\!\! 1) ^3 \pi ^3} \!
\int _{0} ^{\frac{\pi }{2}}
\!\!\! d\theta
\frac{1}{1 \!\!+\!\! \tan ^2 \theta }
\!\!=\!\!
\frac{7}{\pi ^2}\zeta (3)
},
\end{array}
\label{IntegralCalofA2}
\eeqa\\[-4pt]
where we have used
$\sum _{m = 1}^{\infty }
(2m \!\!-\!\! 1)^{-3}
\!=\! 
(7/8) \!\cdot\! \zeta (3)$,
and
$\zeta (3) 
\!=\!
\pi ^3 \!/ 25.79436
$
\cite{GraRyz.63}.\\[1pt]

To get a finite value of
$\sum _p \! A_{p}$,
expanding
(\ref{CalofA})
around
$
y_{T} ^{(\pm)}
\widetilde{x}_{T} ,
$
(\ref{CalofA})
is boldly
approximated as\\[-14pt]
\beqa
\begin{array}{c}
{\displaystyle
\frac{\sum _p A_p}{2 N(0)}
}
\simeq
{\displaystyle
\int _{0} ^{y_T ^{(\pm)}}
\!\! dy
\frac{1}{y}
\tanh y
-
\frac{3}{2}
\left( \!
y_T ^{(\pm)}
\widetilde{x}_T \!
\right) ^2 \!\!
\int _{0} ^{y_T ^{(\pm)}}
\!\! dy
\left\{
\frac
{1}{y ^3}
\tanh y
-
\frac{1}{y ^2}
\mbox{sech} ^2 y
\right\}
}.
\end{array}
\label{CalofA2}
\eeqa
In a similar way 
we also get a roughly approximated integral form for 
(\ref{CalofC}) as\\[-14pt]
\beqa
\begin{array}{c}
{\displaystyle
\frac{\widetilde{\Delta }_T ^2 \!\sum _p \! C_p}{2 N(0)}
}
\simeq
{\displaystyle
\left( \!
y_T ^{(\pm)}
\widetilde{x}_T \!
\right) ^2 \!\!
\int _{0} ^{y_T ^{(\pm)}}
\!\! dy
\left\{
\frac
{1}{y ^3}
\tanh y
-
\frac{1}{y ^2}
\mbox{sech} ^2 y
\right\}
} .
\end{array}
\label{CalofC2}
\eeqa
Integrations of
(\ref{CalofA2}) 
and 
(\ref{CalofC2})
are easily made by using the integration formula
(\ref{integformula})
if we take
the upper-value
$y_T ^{(\pm)}$
to be infinite.

\vspace{1.5cm}

%%%%%%%%%%%%%%
%                                    %
%  Acknowledgements   %
%                                    %
%%%%%%%%%%%%%%

\begin{center}
{\bf Acknowledgements}
\end{center}
S. N. would like to
express his sincere thanks to
Professor Manuel Fiolhais for kind and
warm hospitality extended to
him at the Centro de F\'\i sica Computacional,
Universidade de Coimbra, Portugal.
This work was supported by FCT (Portugal) under the project
CERN/FP/83505/2008.

\newpage

%%%%%%%%%
%                     %
%  References  %
%                     %
%%%%%%%%%

\end{document}